\documentclass[fleqn,usenatbib]{mnras}

\usepackage[T1]{fontenc}

\DeclareRobustCommand{\VAN}[3]{#2}
\let\VANthebibliography\thebibliography
\def\thebibliography{\DeclareRobustCommand{\VAN}[3]{##3}\VANthebibliography}

\usepackage{graphicx}	
\usepackage{amsmath}	
\usepackage{amssymb}	
\usepackage[normalem]{ulem} 
\usepackage{appendix}

\usepackage{lineno}

\def\gtorder{\mathrel{\raise.3ex\hbox{$>$}\mkern-14mu
             \lower0.6ex\hbox{$\sim$}}}
\def\ltorder{\mathrel{\raise.3ex\hbox{$<$}\mkern-14mu
             \lower0.6ex\hbox{$\sim$}}}

\title[DART]{UV to near-IR observations of the DART-Dimorphos collision}

\author[Ofek et al.]{Eran O. Ofek$^{1}$\thanks{E-mail: eran.ofek@weizmann.ac.il},
Doron Kushnir$^{1}$,
David Polishook$^{2}$,
Eli Waxman$^{1}$,
Aaron Tohuvavohu$^{3}$,
\newauthor
Sagi Ben-Ami$^{1}$,
Boaz Katz$^{1}$,
Orly Gnat$^{4}$,
Nora L. Strotjohann$^{1}$,
Enrico Segre$^{2}$,
Arie Blumenzweig$^{1}$,
\newauthor
Yahel Sofer-Rimalt$^{1}$,
Ofer Yaron$^{1}$,
Avishay Gal-Yam$^{1}$,
Yossi Shvartzvald$^{1}$,
Michael Engel$^{1}$,
\newauthor
S. Bradley Cenko$^{5,6}$,
Ofir Hershko$^{2}$\\
$^{1}$Department of Particle Physics and Astrophysics, Weizmann Institute of Science, 76100 Rehovot, Israel.\\
$^{2}$Department of Physics Core Facilities, Weizmann Institute of Science, Herzl St 234, Rehovot 7610001, Israel\\
$^{3}$Department of Astronomy and Astrophysics, University of Toronto, Toronto, ON, Canada\\
$^{4}$Racah Institute of Physics, The Hebrew University, Jerusalem 91904, Israel\\
$^{5}$Astrophysics Science Division, NASA Goddard Space Flight Center, MC 661, Greenbelt, MD 20771, USA\\
$^{6}$Joint Space-Science Institute, University of Maryland, College Park, MD 20742, USA }

\pubyear{2021}

\begin{document}

\label{firstpage}
\pagerange{\pageref{firstpage}--\pageref{lastpage}}
\maketitle

\begin{abstract}

The impact of the Double Asteroid Redirection Test (DART) spacecraft with Dimorphos allows us to study asteroid collision physics,
including momentum transfer, the ejecta properties, and the visibility of such events in the Solar System.
We report observations of the DART impact in the ultraviolet (UV), visible light, and near-infrared (IR) wavelengths.
The observations support the existence of at least two separate components of the ejecta:
a fast and a slow component.
The fast-ejecta component is composed of a gaseous phase, moving at about $1.6$\,km\,s$^{-1}$ with a mass of $\lesssim10^{4}$\,kg.
The fast ejecta
is detected in the UV and visible light,
but not in the near-IR $z$-band observations.
Fitting a simplified optical thickness model to these observations allows us to constrain some
of the properties of the fast ejecta, including its scattering efficiency and the opacity of the gas.
The slow ejecta component is moving at typical velocities of up to about 10\,m\,s$^{-1}$. It is composed of 
micrometer-size particles, that
have a scattering efficiency, at the direction of the observer, of the order of $10^{-3}$ and a total mass of $\sim10^{6}$\,kg.
The larger particles in the slow ejecta, whose size is bound to be in the range between $\sim1$\,mm to $\sim1$\,m,
likely have a scattering efficiency larger than that of the pre-impact Didymos system.

\end{abstract}

\begin{keywords}
minor planets: individual: Didymos ---
minor planets: individual: Dimorphos ---
\end{keywords}

\section{Introduction}
\label{sec:intro}

The Double Asteroid Redirection Test (DART; \citealt{Rivkin+2021PSJ_DART_MissionRequirments,Cheng+2018P&SS_DART_mission_Didymos_DeflectionMission}) examined
planetary defense strategies via the reaction of the near-Earth asteroid
Dimorphos, a satellite of Didymos (65803), to a spacecraft impact (\citealt{Daly+2023Nature_DART_KineticImpact,Cheng+2023Nature_DART_MomentumTransfer, Graykowski+2023Nature_DART_LightCurve}).
The main goal of the DART mission was to measure the momentum gain due to the impact (\citealt{Cheng+2018P&SS_DART_mission_Didymos_DeflectionMission, Cheng+2023Nature_DART_MomentumTransfer, Thomas+2023Nature_DART_OrbitalPeriodChange, Gudebski2023RNAAS_DART_Impact_Anomalous_Orbital_Decay}).
Indeed,
measurements of the orbital period change (\citealt{Thomas+2023Nature_DART_OrbitalPeriodChange,Gudebski2023RNAAS_DART_Impact_Anomalous_Orbital_Decay}),
of the Didymos-Dimorphos system, estimated the momentum gain by Dimorphos due to the DART impact to be about 4 (\citealt{Cheng+2023Nature_DART_MomentumTransfer}),
which is consistent with the pre-impact predictions
(e.g., $1$--$5$; \citealt{Fahnestock+2022PSJ_DART_Predictions}).
DART observations already revealed several intriguing facts,
including ejecta with a large spread of velocities, 
the presence of micro-meter-sized particles as well as meter-size boulders in the ejecta,
and a high-velocity gaseous phase in the ejecta
(e.g., \citealt{Shestakova+2023Icar_DART_Impact_Spectroscopy_EmissionLines, Roth+2023arXiv_DART_Impact_ALMA_Observations, Moreno+2023PSJ_DART_Impact_Ejecta_Observations_and_Models, Jewitt+2023ApJ_DART_Impact_DimorphosBoulderSwarm}).
Detailed observations of the velocity and particle-size distribution
in the ejecta can be used to calibrate the models and assess the target properties (e.g., \citealt{Housen+1983JGR_CrateringScalingLaws_DimensionalAnalysis,Fahnestock+2022PSJ_DART_Predictions,Sanchez+2022arXiv_DART_Didymos_ImpactSeismicWaves, Kumamoto+2022arXiv_DART_Didymos_PredictionSimulations}).

Furthermore, the DART mission gives us some insight into the physics and visibility of asteroid collisions (e.g., \citealt{Farinella+Davis1992Icar_AsteroidsCollisionsVelocity, Holsapple+Housen2019P&SS_AsteroidsCollisions_Review, Holsapple2022P&SS_MainBeltAsteroids_CollisionsHistory_Ejecta_Qstar}, Ofek et al., in prep.).
Asteroid collisions were reported in the past (e.g., \citealt{Jewitt+2010Natur_AsteroidCollision_P2010A2, Jewitt+2011ApJ_596_Scheila_AsteroidCollision_HST_observations, Haisch+1991_FlareStars_Review, Masateru+2011ApJ_596_Scheila_AsteroidCollision_Observations, Moreno+2011ApJ_596_Sceila_AsteroidCollision_Observations}), but in all cases were likely detected many days or weeks after the impact.
These collisions are also likely responsible for the production of some of the zodiacal dust (e.g., \citealt{Gustafson+1994AREPS_PhysicsZodiacalLight_Review}).

Here, we report on ultraviolet (UV), visible, and near-infrared (NIR) observations of the Dimorphos-DART impact.
We show that the ejecta was composed of two main components.
A slow component expanding at a velocities of the order of a m\,s$^{-1}$ 
and a fast component moving at 1,600\,m\,s$^{-1}$.
The brightening of the event after the collision is interpreted as a transition from an optically thick to an optically thin phase.
This, as well as the surface brightness evolution, allows us to measure the scattering and opacity of the emitting gas in the fast ejecta.
Furthermore, these observations may have implications for the visibility of asteroid collisions in the main belt, which in turn may be able to constrain the number of meter-sized asteroids (Ofek et al., in prep.).

The structure of this paper is as follows.
In \S\ref{sec:obs} we describe the observations, while in 
\S\ref{sec:EffectiveSurfaceArea}, we define the surface area units that we use, and in \S\ref{sec:overview}, we provide an overview of the ejecta.
In \S\ref{sec:fast}, we discuss the fast-ejecta component, while in \S\ref{sec:slow} the slow-ejecta component.
Finally, we conclude in \S\ref{sec:conclusion}.

\section{Observations}
\label{sec:obs}

We obtained observations of the DART impact using
the Large Array Survey Telescope (LAST; \S\ref{sec:LAST}), the {\it Swift}-UVOT space telescope (\S\ref{sec:Swift})
and the Rich 28-inch telescope (C28) at the Wise Observatory (\S\ref{sec:C28}).
The Observations are summarized in Table~\ref{tab:Obs}.
All the magnitudes in this paper are in the Vega system.

\begin{table}
\centering
\begin{tabular}{llll}
\hline
Telescope  & Filter & Observing window & Number of images\\
        &           & (2022 UTC)       & \\
\hline
LAST       & clear  & 09-26/27 23:12--02:38  & 2481 \\
           & clear  & 09-27/28  23:02--02:37 & 2350 \\
           & clear & 09-28/29 22:30--02:30 & 2229 \\
           & clear & 09-29/30 22:32--02:40 & 2474 \\
           & clear & 10-01/02 22:12--01:47 & 1840 \\
           & clear & 10-03/04 22:39--02:30 & 437 \\
           & clear & 10-04/05 22:34--02:22 & 196 \\
C28        & $z$   & 09-26/27 23:00--02:30 & 645 \\
UVOT       & $UVW2$ & 09-26 23:08--23:20 & Photon counting \\
\hline
\end{tabular}
\caption{\label{tab:Obs} 
Summary of observations.}
\end{table}

\subsection{The Large Array Survey Telescope}
\label{sec:LAST}

The Large Array Survey Telescope (LAST) is an under-construction array of 48, 28-cm f/2.2 telescopes (\citealt{Ofek+BenAmi2020_Grasp_SkySurvrys_CostEffectivness,BenAmi+2023PASP_LAST_Science}).
At the time of DART impact, we used six out of 12 telescopes installed at that time\footnote{Currently 32 telescopes are installed.}.
Each telescope provides a field of view of 7.4\,deg$^{2}$, and pixel scale of $1.25''$\,pix$^{-1}$.
The LAST observations were taken without a filter. The LAST response is similar to that of the GAIA $B_{\rm p}$ band, but extending more into the red
(from 4000\,\AA~to 8000\,\AA~and peaking at 4800\,\AA), and it is calibrated to this band.
During the impact, we observed with 5\,s integrations and we switched to 20\,s, after a few tens of minutes.

The LAST images were dark subtracted and flat corrected
using tools available in the LAST data pipeline (\citealt{Ofek+2023ASP_LAST_PipelineI, Ofek2014_MAAT, Ofek2019_Astrometry_Code, Soumagnac+Ofek2018_catsHTM}).
The code used for the data reduction is available online via GitHub\footnote{https://github.com/EranOfek/AstroPack}.
The Didymos light curve was obtained using a specialized version of the pipeline in which the photometry is performed on the predicted astrometric position of the asteroid.
Given the extended nature of the target and the high airmass of observations, we have used several large photometric apertures of $5''$, $7.5''$, $10''$, and $12.5''$.
The LAST photometry was calibrated against the $B_{\rm p}$ magnitude of a single GAIA star,
on each night.
To minimize the color term in the calibration, we choose stars with $B_{\rm p}-R_{\rm p}\approx0.8$.
In Table~\ref{tab:LASTphot}, we list all 12,007 photometric measurements collected by the LAST system.
The $B_{\rm p}$ light curve of the event is shown in Figure~\ref{fig:DART_LAST_mag_logt_apersPhases},
where the flux is presented in units of the effective surface area defined in \S\ref{sec:EffectiveSurfaceArea}.
\begin{table*}
\centering
\begin{tabular}{lllllllll}
\hline
 JD       &     Mag(5) & Mag(7.5) & Mag(10) & Mag(12.5) &Err(5) &Err(7.5) &Err(10) &Err(12.5)\\  
 (day)    &    (mag)   &  (mag)   & (mag)   & (mag)     & (mag) & (mag)   & (mag)  & (mag)   \\
\hline\hline 
2459849.4671696  &   14.674  &14.701  &14.669  &14.723  &  0.026  &0.025  &0.024  &0.024  \\       
2459849.4672276  &   14.706  &14.741  &14.743  &14.779  &  0.026  &0.025  &0.025  &0.025  \\     
2459849.4672855  &   14.581  &14.676  &14.719  &14.807  &  0.025  &0.025  &0.025  &0.026  \\       
2459849.4673433  &   14.695  &14.712  &14.741  &14.667  &  0.026  &0.025  &0.025  &0.024  \\       
2459849.4674013  &   14.608  &14.573  &14.538  &14.465  &  0.025  &0.023  &0.022  &0.021  \\       
\hline
\end{tabular}
\caption{\label{tab:LASTphot} 
List of all 12,007 observations of the Didymos system collected using LAST. There are four magnitude and error columns corresponding to
the aperture photometry around the predicted Didymos position in apertures with radii of $5''$, $7.5''$, $10''$, and $12.5''$.
The full Table is available in the electronic version of this paper. Here we list the first five entries.}
\end{table*}

\begin{figure}
\centering
\includegraphics[width=\linewidth]{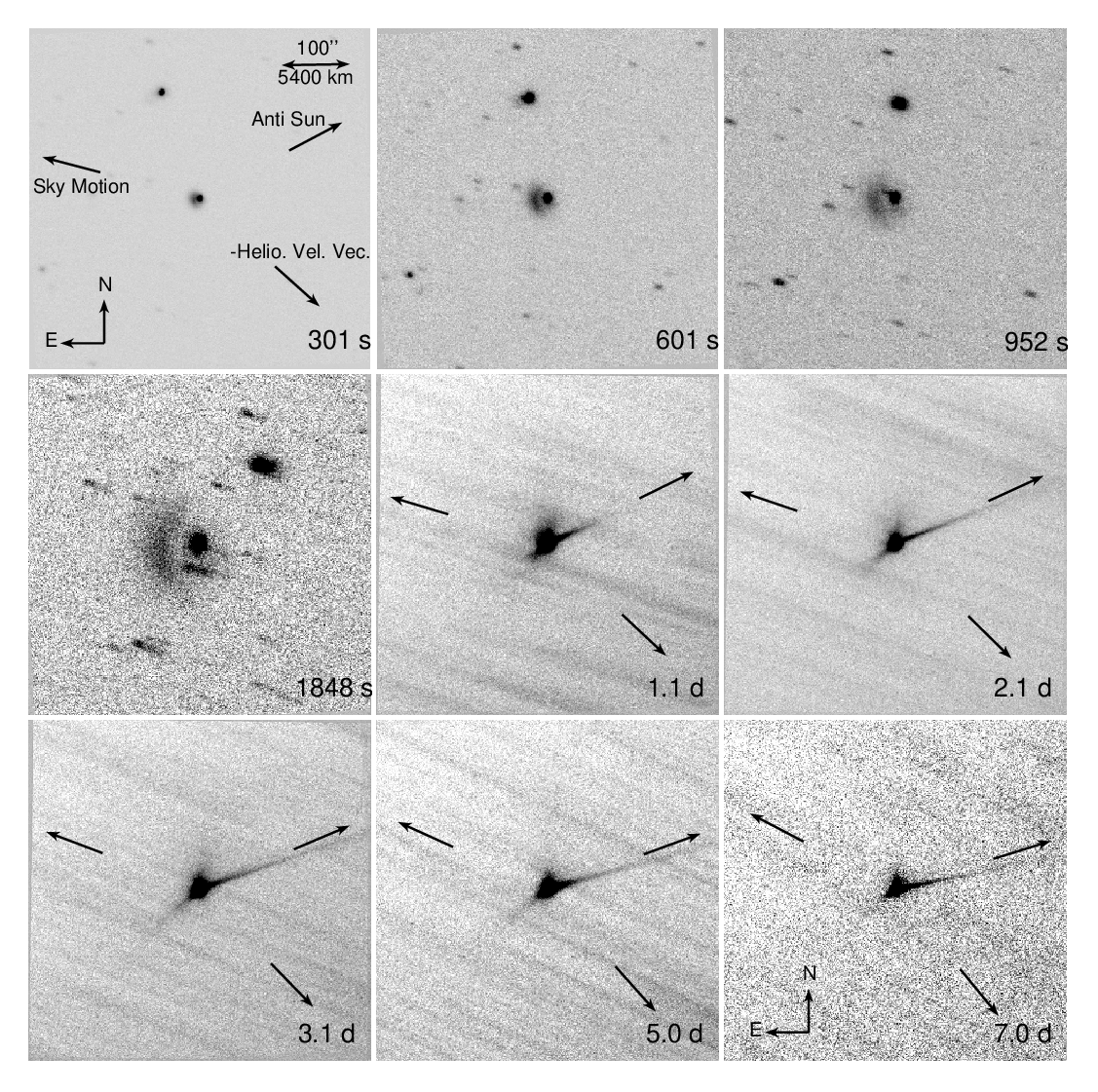}
\caption{LAST $B_{\rm p}$ images of Didymos at selected times after the DART impact obtained by coadding (\citealt{Zackay+2017_CoadditionI}) a large number of images from different LAST telescopes. Also shown are the sky motion, antisolar direction, and the negative of the Heliocentric velocity vector of the asteroid.}
\label{fig:DART_Images}
\end{figure}

\begin{figure}
\centering
\includegraphics[width=\linewidth]{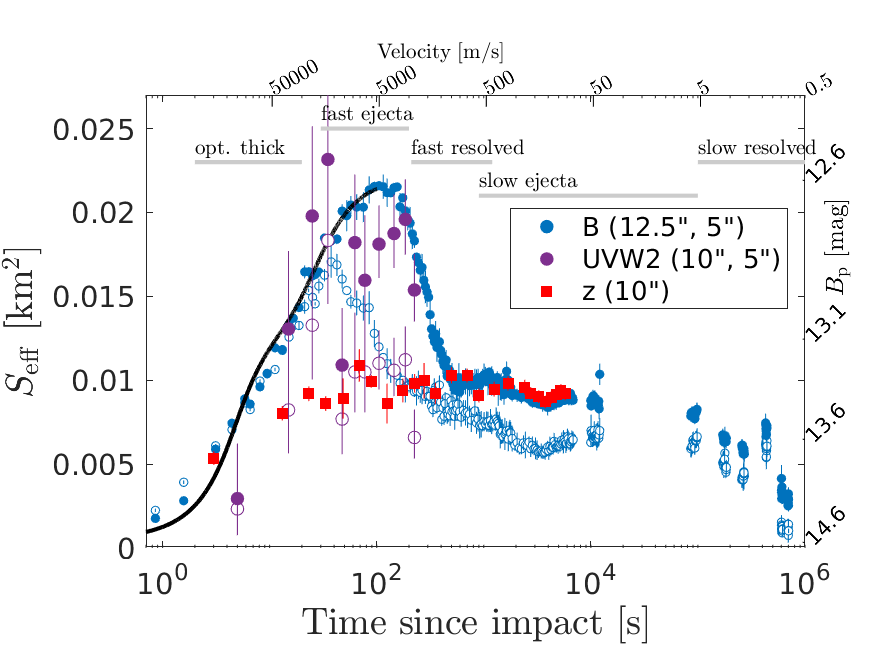}
\caption{The Didymos system excess flux
in units of effective area for scattering 100\% of the solar radiation (\S\ref{sec:EffectiveSurfaceArea}).
Time is measured relative to the light-time corrected impact time (0.2\,s uncertainty; UTC 2022-09-26 23:15:01.9; LAST topocentric time).
The filled (empty) purple points show the binned {\it Swift}-UVOT UVW2-band excess flux for a $10''$ ($5''$)-radius aperture, while the blue filled (empty) points show the binned $B_{\rm p}$-band flux for $12.5''$ ($5''$)-radius apertures.
The filled red boxes present the binned $z$-band light curve in the $10''$-radius aperture.
The corresponding $B_{\rm p}$ magnitude is shown on the right axis.
The $z$-band magnitude can be obtained by subtracting 1\,mag,
while for the $UVW2$ add $4.5$\,mag.
The top axis shows the velocity for which the ejecta would expand beyond the $10''$ aperture at that time.
This velocity estimate is an approximation as it does~not take into account the point spread function and ejecta geometry.
The solid black curve shows the best fit (to the non-binned data) of our optical thickness model for the fast and slow ejecta (see also Figure~\ref{fig:fitRiseLC_t0_slow_fast}).
To mimic the 5\,s exposures, the optical thickness model was convolved with a 5\,s top-hat function. All error bars show 1-$\sigma$ errors.
The gray lines at the top show different phases of the light curve.}
\label{fig:DART_LAST_mag_logt_apersPhases}
\end{figure}

The post-impact images of the Didymos system are shown in Figure~\ref{fig:DART_Images}.
These images are based on the coaddition of multiple LAST images collected using several telescopes.
The images were first registered to the predicted position of the Didymos asteroid, background subtracted, and flux matched (by the image photometric zero points),
and then coadded using a median sigma-clipped coaddition. 
Table~\ref{tab:coadd} summarizes the properties of the coadded images.

\begin{table}
\centering
\begin{tabular}{lll}
\hline
Mid time & Time range & Exposure time \\
(s)      & (s)        & (s) \\
\hline\hline
301.3 &119.3 & 480\\
601.4 &119.3 & 480\\
952.3 &187.8 & 770\\
1849 &372    & 1415\\
92268 & 12903 &47260\\
177132 & 14389& 43365\\
264763 & 14839& 50140\\
434459 & 12876& 36880\\
609478 & 13819& 12900\\
\hline
\end{tabular}
\caption{\label{tab:coadd} 
Properties of coadd images (Figure~2). Mid-time in seconds since impact,
time range and total exposure time of all the coadded images.}
\end{table}

\subsection{Swift-UVOT}
\label{sec:Swift}

We also observed the event using
the {\it Swift}-UVOT (\citealt{Gehrels+2004_Swift, Roming+2005SSRv_Swift_UVOT}), in event mode (11\,ms time resolution) and the $UVW2$ filter (1600\,\AA\,to 3000\,\AA~and peaking at about 2000\,\AA).
The {\it Swift}-UVOT is a 30-cm telescope,
with 0.5"/pix, equipped with intensified CCD that can be operated
in photon counting mode.

The {\it Swift} UVW2 observations, presented in Figure~\ref{fig:DART_LAST_mag_logt_apersPhases} and Table~\ref{tab:Swift},
detected Didymos prior to the impact at $UVW2$ Vega magnitude of $19.0\pm0.1$.
The UV excess flux peaked at $16.4_{-0.2}^{+0.3}$, about 35\,s after the impact.


\begin{table*}
\centering
\begin{tabular}{lllllll}
\hline
Mid Time & Mag(5)     & Low Err(5) & High Err(5) & Mag(10)     & Low Err(10) & High Err(10)\\
(s)      & (mag)   & (mag)   & (mag)  & (mag)   & (mag)   & (mag) \\
\hline
\hline
   5.0 & 18.15 & 0.41 & 0.93    & 18.00 & 0.43 & 0.75\\
  15.0 & 17.19 & 0.28 & 0.47    & 16.76 & 0.25 & 0.34\\
  25.0 & 16.75 & 0.23 & 0.36    & 16.36 & 0.21 & 0.27\\
  35.0 & 16.43 & 0.20 & 0.29    & 16.20 & 0.19 & 0.25\\
  47.5 & 17.25 & 0.24 & 0.37    & 16.93 & 0.22 & 0.29\\
  62.5 & 16.97 & 0.21 & 0.31    & 16.44 & 0.18 & 0.22\\
  77.5 & 16.97 & 0.21 & 0.31    & 16.57 & 0.19 & 0.24\\
 105.0 & 16.92 & 0.13 & 0.17    & 16.44 & 0.11 & 0.13\\
 145.0 & 16.96 & 0.13 & 0.17    & 16.41 & 0.11 & 0.12\\
 185.0 & 16.91 & 0.13 & 0.16    & 16.37 & 0.11 & 0.12\\
 225.0 & 17.38 & 0.16 & 0.21    & 16.61 & 0.12 & 0.14\\
\hline
\end{tabular}
\caption{\label{tab:Swift} 
Summary of {\it Swift}-UVOT observations. The observations were obtained using the {\it UVW2} filter in photon counting mode, and the number of photons in each time bin was converted to Vega magnitudes. The errors were calculated assuming Poisson statistics (\citealt{Gehrels1986_PoissonCI}). 
The flux was converted to magnitudes using the instrument zero point.
The time is measured relative to the light-travel time corrected impact time.
The bin sizes are adaptive. The first bin starts in 0 seconds, and the mid-time indicates the middle of each time bin.
The measurements are based on aperture photometry with $5''$ and $10''$ aperture radii. The pre-impact magnitude of the Didymous system was $19.04$.}
\end{table*}

\subsection{Wise 28-inch telescope}
\label{sec:C28}

We observed using the Jay Baum Rich 28-inch (C28) telescope (\citealt{Brosch+2015ApSS_C28_Telescope_WiseObservatory}) in the Wise Observatory, Mizpe-Ramon, Israel
(about 65\,km North of the LAST system).
This telescope has a pixel scale of $0.83''$\,pix$^{-1}$.
The C28 observations were conducted using a $z$-band filter (8100\,\AA\,to 10000\,\AA).

The C28 observations were calibrated against the SkyMapper catalog (\citealt{Keller+2007PASA_SkyMapper, Wolf+2018PASA_SkyMapper_DR1}).
The pre-impact $z$-band magnitude of the system was $13.7\pm0.2$ (Vega) mag, and it peaked at about 12.4\,mag,
about 100\,s after the impact.
The 600 C28 observations, are listed in Table~\ref{tab:C28} and presented in Figure~\ref{fig:DART_LAST_mag_logt_apersPhases}.

\begin{table}
\centering
\begin{tabular}{lll}
\hline
Time & Mag & Err \\
(s)  & (mag) & (mag) \\
\hline
\hline
  $-1778.5$ &14.02 &0.43\\
  $-1768.3$ &13.97 &0.27\\
  $-1758.2$ &13.19 &0.21\\
  $-1748.0$ &13.53 &0.30\\
  $-1737.9$ &13.03 &0.15\\
\hline
\end{tabular}
\caption{\label{tab:C28} 
Summary of the C28 observations. The observations were obtained using the {\it z} filter.
The measurements are based on aperture photometry with $10''$ aperture radius. We list here the first five measurements, while the full Table is available with the electronic version of this manuscript.}
\end{table}

\section{Surface area units}
\label{sec:EffectiveSurfaceArea}

We display the light curve of the DART impact using units of effective surface area ($S_{\rm eff}$).
The reason is that surface area units have a physical meaning relevant to reflection (the surface area of the reflecting particles) and to emission (the surface area is related to the mass of the particle via the opacity $\kappa$).
For a geometry in which the dimensions of the scattering cloud in the radial and transverse directions
is similar, we can write:
\begin{equation}
    \kappa \sim \frac{S_{\rm eff}}{M},
    \label{eq:kappa}
\end{equation}
where $M$ is the mass of the ejecta.
Here we define the effective surface area $S_{\rm eff}$ such that
\begin{equation}
    f_{\rm x} = f_{\odot, \rm x} N_{\rm p} \frac{d\sigma}{d\Omega}(\theta) \frac{1}{\Delta^{2}} \equiv f_{\odot, \rm x} S_{\rm eff} \frac{1}{\Delta^{2}},
    \label{eq:flux}
\end{equation}
where $f_{\rm x}$ is the observed flux in band $x$, $f_{\odot, \rm x}=L_{\odot, \rm x}/(4\pi r^{2})$ is the solar flux
at band $x$, $L_{\odot, \rm x}$ is the solar luminosity, $r$ and $\Delta$ are the target Heliocentric and Geocentric distances, respectively,
$N_{\rm p}$ is the number of particles and $d\sigma/d\Omega(\theta)$ is the cross section for scattering at angle $\theta$
(with units of sr$^{-1}$).

Using the definition for $S_{\rm eff}$, its relation to magnitude in band $x$ is given by:
\begin{equation}
    S_{\rm eff, tot} = 10^{0.4(M_{\odot,\rm x} - M_{\rm x}) + 2\log_{10}(r\Delta)},
\end{equation}
where $M_{x}$ is the apparent magnitude of the asteroid and $M_{\odot,x}$ is the Geocentric apparent magnitude of the Sun, in band $x$.
For $r$ and $\Delta$ in AU, $S$ is in AU$^{2}$.
To derive the excess surface area of the ejecta, we subtract the Didymos surface area as calculated from the pre-impact magnitude of Didymos ($M_{\rm pre, x}$):
\begin{equation}
    S_{\rm eff} = S_{\rm eff, tot}(M_{\rm x}) - S_{\rm eff, tot}(M_{\rm pre, x}).
\end{equation}

\section{Overview description of the light curve}
\label{sec:overview}

Following the DART impact, we observe two main resolved ejecta components. On time scales of the order of ten minutes after the impact, a spherical-cup-like component is detected moving eastward. This component has an opening angle of about two radians.
We refer to this component as the {\it fast ejecta}.
This ejecta is seen up to about one hour after the impact, when its surface brightness drops below our detection limit (see Figure~\ref{fig:DART_Images}, \ref{fig:EjectaSurfaceBrightness}).
After about a day, we resolve another, seemingly distinct, part of the ejecta we refer to as the {\it slow ejecta}. This component is composed of a fan line component and a tail (see also, e.g., \citealt{Daly+2023Nature_DART_KineticImpact, Li+2023Nature_DART_EjectaActiveAsteroid, Roth+2023arXiv_DART_Impact_ALMA_Observations, Moreno+2023PSJ_DART_Impact_Ejecta_Observations_and_Models}).
Spectroscopic observations of the fast ejecta (\citealt{Shestakova+2023Icar_DART_Impact_Spectroscopy_EmissionLines}) showed that a major part of the light of this component arrives from emission lines of neutral species (\ion{Na}{I} $\lambda$ 5890, 5896\,\AA,~\ion{K}{I} $\lambda$ 7699\,\AA,~\ion{Li}{I} $\lambda$ 6708\,\AA), suggesting that the DART impact released gas.
The origin of this gas is unknown but could be either a trapped gas or a gas created by the impact itself.
For example, energy conservation does~not prohibit the creation of a gaseous phase. Specifically, by equating the kinetic energy to the thermal energy, we can get an upper limit on the temperature of the heated material in the impact:
\begin{equation}
    T \lesssim \frac{m_{\rm A} v_{\rm imp}^{2}}{k_{\rm B}}.
\end{equation}
Here $m_{\rm A}$ is the atom mass, $v_{\rm imp}\cong6,145$\,m\,s$^{-1}$ (\citealt{Cheng+2023Nature_DART_MomentumTransfer}) is the impact velocity, and $k_{\rm B}$ is the Boltzmann constant.
For Sodium atoms, this provides an upper limit of about $10^{5}$\,K.
However, this is likely limited to a small fraction of the ejecta.
Interestingly, such intermittent Sodium emission was also detected in other Solar System objects like Phaethon (3200) and Mercury (e.g., \citealt{Lierle+2023arXiv_HighResSpectroscopy_Mercury_NaI, Zhang+2023PSJ_Phaethon3200_NaI_emission}).

It is known that neutral species can not survive for long periods of time in the radiation field of the Sun.
For example, \cite{Shestakova+2023Icar_DART_Impact_Spectroscopy_EmissionLines} estimate that at distance of 1\,AU from the Sun, \ion{Na}{I} and \ion{K}{I} will become ionized on time scale of 1\,day, while \ion{Li}{I} will become ionized within about 1\,hr.
Therefore, given that the fast-ejecta visible light is dominated by \ion{Na}{I} and \ion{K}{I} emission lines, we estimate that during our fast-ejecta observations, the emission lines' luminosity does~not change considerably. 
This assumption is roughly confirmed by the measurement of the surface brightness evolution (\S\ref{sec:SurfaceBrightnessEvolution}).

Following the impact, the light curve rises in all bands within 100\,s. We attribute this rise to an optical thickness effect,
{\it i.e.}, at early times, when the density of the ejecta is high enough, some of the gas or dust (depending on the component) is being self-obscured.
In Appendix~\ref{app:Model_t0}, we derive a simplified model for the optical thickness of the ejecta.
Following the rise, the UV and blue optical observations show a peak lasting for about 100\,s followed by a drop to an intermediate level. The first drop in the light curve is attributed to the fast ejecta leaving the photometric aperture.
Indeed, the peak of the light curve, as detected in the larger photometric aperture, lasts longer than the peak at the smaller photometric apertures.
Contrary to the UV and visible light, the $z$-band observations rise, but directly to the intermediate level.
This suggests that there are no prominent emission lines in the $z$ band (in the range of 8100\,\AA~to 10000\,\AA).
About one day after the impact, the light curve starts to drop, on a time scale of about one week, from its intermediate level back to the pre-impact luminosity.
This second drop is due to the slow ejecta leaving the photometric aperture.

Finally, at later times, solar radiation sorts the ejecta particles by size, where micro-meter size particles form a tail.

\section{The fast ejecta}
\label{sec:fast}

To measure the velocity of the fast ejecta,
we linearly fitted the time-dependent angular distance
between the asteroid and the 
brightest pixel in the fast ejecta, convolved with the point spread function (PSF) -- i.e., a matched filter detection.
Next, we converted this angular speed to velocity, using the known distance to the Didymos system.
The fast ejecta position angle of motion is 108\,deg, and its projected speed is $v_{\rm ej, fast}=1.6\pm0.2$\,km\,s$^{-1}$.
Since the expected direction of the impact, and hence the ejecta, is about 90\,deg from the observer-Dimorphos line (e.g., \citealt{Rivkin+2021PSJ_DART_MissionRequirments}), this measured velocity is likely close to the actual ejecta velocity.
The largest uncertainty in the velocity is due to the on-sky projection angle,
which here we assume is known to about 10\% accuracy.
This estimate is consistent with the fast ejecta velocity measured by \cite{Shestakova+2023Icar_DART_Impact_Spectroscopy_EmissionLines}.
However, it differs from the value given in \cite{Graykowski+2023Nature_DART_LightCurve} (of $970\pm50$\,m\,s$^{-1}$). These authors estimate the velocity from the light curve – i.e., from the time it takes the ejecta light to get out of the photometric aperture.
However, such an estimate depends on the exact ejecta geometry and seeing.

We can use this velocity, along with
energy conservation, to put an upper limit of the mass of the fast ejecta:
\begin{equation}
    m_{\rm ej, fast} \lesssim m_{\rm imp} \Big(\frac{v_{\rm imp}}{v_{\rm ej, fast}}\Big)^{2}.
    \label{eq:EnergyCons}
\end{equation}
Here, $m_{\rm imp}=579.4$\,kg is the impactor mass, and $v_{\rm imp}=6.145$\,km\,s$^{-1}$ is the impactor velocity (\citealt{Cheng+2023Nature_DART_MomentumTransfer}).
This translates to an upper limit of $m_{\rm ej, fast}\lesssim10^{4}$\,kg.
We note that if the mass of the fast ejecta is close to this limit, then the measured momentum gain of the impact (\citealt{Cheng+2023Nature_DART_MomentumTransfer}) is dominated by this fast ejecta.
However, estimating the mass of the fast ejecta requires a non-steady-state radiative model.

We estimate the effective area,
defined in \S\ref{sec:EffectiveSurfaceArea}, of the fast-ejecta component, by reading the total surface area of the peak in Figure~\ref{fig:DART_LAST_mag_logt_apersPhases} and subtracting the surface area of the plateau region (slow ejecta, at around an hour after the impact). 
We find that the effective surface area of the fast ejecta is about
$0.0125$\,km$^{2}$.
This value suffers from several uncertainties that are not straightforward to estimate, including what is the fraction of slow-ejecta surface brightness during the fast-ejecta peak, which is still optically thick, and the existence of additional ejecta components.
This estimate is also consistent with a direct measurement of the total flux of the resolved fast ejecta
we obtained by integrating the light of the fast ejecta in the LAST images taken about
30\,min after the impact.
The effective surface area (or flux) ratio of the fast-ejecta component between the different bands ($UV$, $B_{\rm p}$ and $z$) are:
$0.3<S_{\rm eff, fast, UV}/S_{\rm eff, fast, B}<3$, and $S_{\rm eff, fast, B}/S_{\rm eff, fast, z}>4$.

We note that prior to the finding that the fast-ejecta light is mainly due to emission lines (\citealt{Shestakova+2023Icar_DART_Impact_Spectroscopy_EmissionLines}), we interpreted these observations as Mie-scattering on sub-micron particles. Such a model allows the estimation of the fast-ejecta mass (including the absorbing/unseen particles) and even the dielectric properties of the particles.
However, given the observations of emission lines, this interpretation can be ruled out (see Appendix~\ref{sec:Mie}).

Finally, using Equation~\ref{eq:kappa}, and the upper limit on the mass of the fast ejecta, we can put a lower limit on the gas effective opacity, of $\kappa\gtrsim10^{4}$\,cm$^{2}$\,g$^{-1}$.

\subsection{The rising light curve}
\label{sec:FitT0}

We interpret the rise of the light curve as a transition from an optically thick to an optically thin phase.
Evidence for the correctness of this interpretation was seen directly in the case of the {\it DeepImpact} collision with Temple\,1 where the shadow of the ejecta is directly seen (\citealt{AHearn+2005Sci_DeepImpact_Tempel1_Comet_ObservationsReview, Harker+2005Sci_DeepImpact_Tempel1_Dust_Collision, Meech+2005Sci_DeepImpact_Tempel1_Collision_Observations}).

In Appendix~\ref{app:Model_t0}, we derive a simplified model for the optical thickness as a function of time, as well as the surface brightness of the event.
This model assumes a single scattering of photons before they escape from the ejecta (i.e., the cross-section for absorption is larger than the cross-section for scattering).
For line emission, this means that a UV photon is not exciting more than one atom.
In this Appendix, we also discuss the caveats in this model.
The ejecta geometry is assumed to be a spherical shell section
with a half-opening angle $\gamma$, and a width over distance ratio of $\delta$. 
The time scale for the optically thick to optically thin transition is
\begin{equation}
    t_{0} = \sqrt{\frac{9 m_{\rm ej} \sigma_{\rm T} }{8\pi^2 \rho r_{\rm p}^3  v_{\rm ej}^{2} (1-\cos(\gamma))} }.
    \label{eq:t0_mono}
\end{equation}
Here $\sigma_{\rm T}$ is the total (absorption and scattering) cross-section of each particle, $\rho$ is the particle density, and $r_{\rm p}$ is its radius.
For gas, one can replace $\rho$ and $r_{\rm p}$ with the particle mass (i.e., $m_{\rm p}=(4/3)\rho r_{\rm p}^{3}$.

\begin{figure}
\centering
\includegraphics[width=\linewidth]{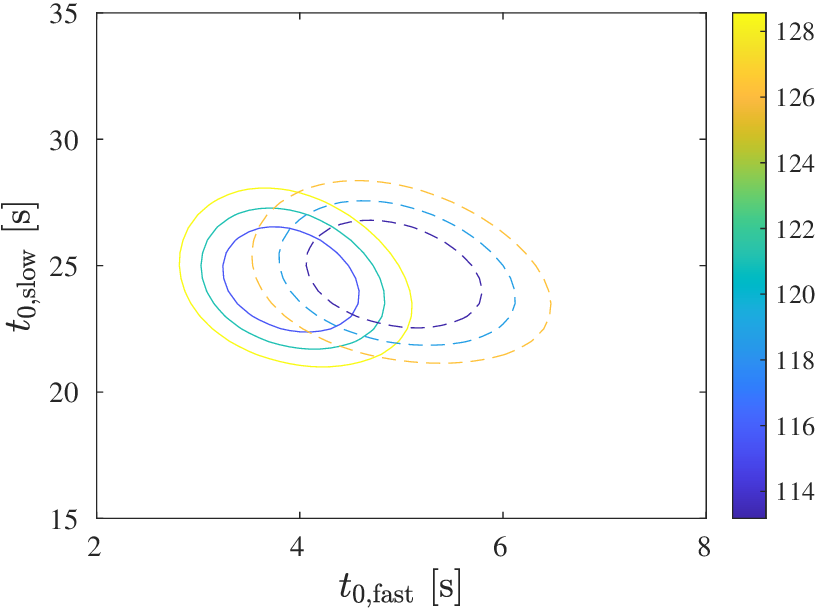}
\caption{The $\chi^{2}$ surface of the best fit $t_{0,\rm fast}$ and $t_{0,\rm slow}$ to the rising light curve (first 100\,s). The solid contours are for fast ejecta with $\gamma_{\rm fast}=1$, $\delta_{\rm fast}=0.5$,
and slow ejecta with $\gamma_{\rm slow}=\pi/2$, $\delta_{\rm slow}=1$,
while the dashed contours are for fast ejecta with
$\gamma_{\rm fast}=1$, $\delta_{\rm fast}=0.2$,
and slow ejecta with $\gamma_{\rm slow}=1.5$, $\delta_{\rm slow}=1$.}
\label{fig:fitRiseLC_chi2_t0_slow_fast}
\end{figure}

We fitted the first 100\,s after impact, of the Didymos $B_{\rm p}$ light curve
with the spherical cup optical thickness model.
We use data obtained in the $12.5''$-radius aperture,
and we assumed $S_{\rm fast, B}=0.0125$\,km$^{2}$, $\gamma_{\rm fast}=1$, and $\delta_{\rm fast}=0.5$ for the fast ejecta,
and $S_{\rm slow,B}=0.0095$\,km$^{2}$, $\gamma_{\rm slow}=\pi/2$, and $\delta_{\rm slow}=1$ for the slow ejecta.
In order to calculate the $\chi^{2}$ between the observations and the model,
we convolved the model with a top-hat function with 5\,s width (to mimic the 5\,s exposure time).
This allows us to probe $t_{0}$, which is shorter than our time resolution.
The $\chi^{2}$ as a function of $t_{0, {\rm fast}}$ and $t_{0, {\rm slow}}$ is presented in Figure~\ref{fig:fitRiseLC_chi2_t0_slow_fast}.
The contours show the 1, 2, and 3-$\sigma$ confidence levels.
To show the sensitivity to other values of $\gamma$ and $\delta$
we also show (in dashed contours) the $\chi^{2}$ contours
for $\gamma_{\rm fast}=1$ and $\delta_{\rm fast}=0.2$ for the fast ejecta,
and $\gamma_{\rm slow}=1.5$ and $\delta_{\rm slow}=1$ for the slow ejecta.
We get best fit values of $t_{0, {\rm fast}}=3.9\pm0.5$\,s and $t_{0,{\rm slow}}=24.5\pm2$\,s (with best fit $\chi^{2}/dof=108/92$).
The errors were calculated assuming six degrees of freedom (to mimic all the free parameters in the problem).
\begin{figure}
\centering
\includegraphics[width=\linewidth]{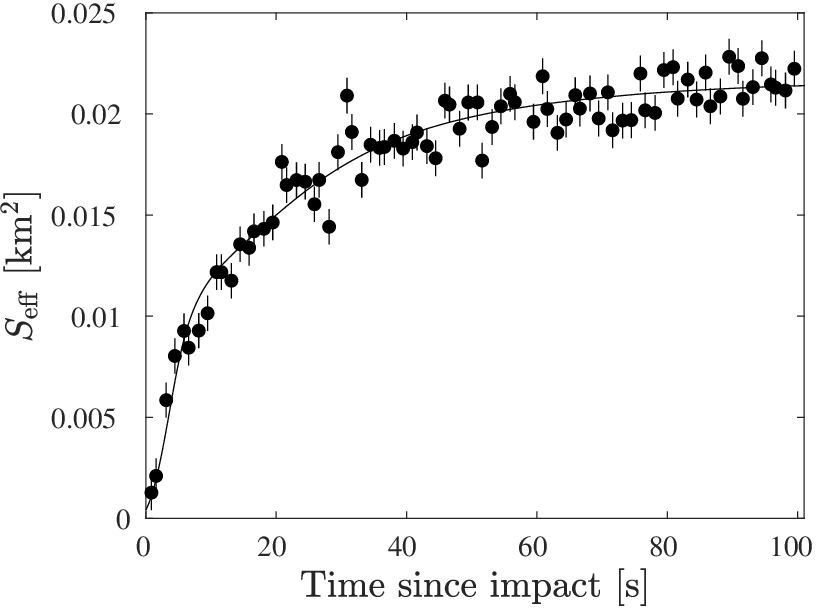}
\caption{The rising (unbinned) light curve (first 100\,s) obtained using LAST in the $B_{\rm p}$ band and photometric aperture radius of $12.5''$. Errors represent 1-$\sigma$ uncertainties. The black line shows the best-fit rise model using $t_{0,\rm fast}=3.9$\,s and $t_{0,\rm slow}=24.5$\,s.}
\label{fig:fitRiseLC_t0_slow_fast}
\end{figure}

\subsection{Surface brightness evolution}
\label{sec:SurfaceBrightnessEvolution}

For constant emission or reflection from particles expanding at a constant velocity, the surface brightness evolves like $t^{-2}$. Therefore, by measuring the surface brightness of the fast ejecta, we can extrapolate its brightness at early times and find the surface brightness at which the fast ejecta becomes optically thin (i.e., $t_{0, \rm fast} \approx 3.9$,s).
This surface brightness is equivalent to the
efficiency in which UV Sunlight excites the atoms that emit
their light in the visible band.
I.e., the 
scattering efficiency ($d\sigma/d\Omega(\theta)/\sigma_{\rm T}$)
of the fast ejecta.
Here $d\sigma/d\Omega(\theta)$ is the cross section for scattering at direction $\theta$ ($53.3$\,deg in our case),
and $\sigma_{\rm T}$ is the total cross section for absorption and scattering.

We measure the maximum surface brightness as a function
of time along the line connecting Didymos with the fast ejecta trajectory
(i.e., position angle of 108\,deg).
After converting these flux measurements to surface brightness,
this provides a measurement of $I_{\rm x, max}$ (the maximum surface brightness, defined in \S\ref{app:Model_t0}).
Assuming $\gamma=1$ and $\delta=0.5$, the ratio
between the maximum surface brightness and the mean surface brightness is about $1.92$ (derived in \S\ref{sec:SurfB}).

\begin{figure}
\centering
\includegraphics[width=\linewidth]{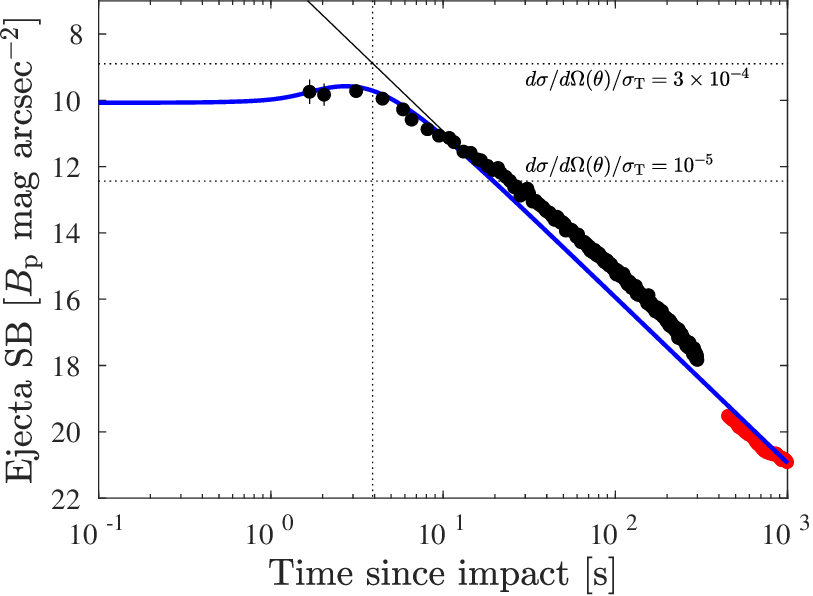}
\caption{The measured $B_{\rm p}$ surface brightness [magnitude\,arcsec$^{-2}$] of the brightest pixel in the fast ejecta (along the motion direction) as a function of time (red circles).
The black circles show the surface brightness as estimated from the photometry divided by $(v_{\rm ej, fast}t)^{2}$,
while the blue line shows the corresponding optical thickness model (see text for details).
The errors of the black dots include the propagated uncertainty due to the pre-impact magnitude of the system and the uncertainty at the time of impact.
The vertical dotted line shows the optical thickness time scale of the fast ejecta (i.e., $t_{0}=3.9$\,s).
The vertical dotted line crosses the surface brightness extrapolation (black line) at $d\sigma/d\Omega(\theta)/\sigma_{\rm T}=2.6\times10^{-4}$.
The dotted horizontal lines show the surface brightness levels
for $d\sigma/d\Omega(\theta)/\sigma_{\rm T}$ of
$10^{-5}$, and $2.6\times10^{-4}$.
\label{fig:EjectaSurfaceBrightness}}
\end{figure}

In Figure~\ref{fig:EjectaSurfaceBrightness}, we show the surface brightness (in magnitude per arcsec$^{2}$) as a function of time. The red points show the measured surface brightness from the resolved ejecta. The power-law index of the red points deviates from $2$ because, presumably, at early times, given our poor spatial resolution, the fast ejecta is not fully resolved (in the motion direction). Therefore, when extrapolating the surface brightness to early times (black line) we anchor the extrapolated line to the latest measurements at $t\approx1000$\,s.
The black dots show our early $12.5''$-radius aperture photometry converted to surface brightness by extrapolating the area of the ejecta by $(v_{\rm ej}(t+dt))^{2}$. Since at the first few seconds, the flux evolves like $t^{2}$, we added $dt$, which is $1/4$ of the exposure time. The blue line shows the best fit optical thickness model for the fast ejecta component (fitted in \S\ref{sec:FitT0} and shown in Figure~\ref{fig:fitRiseLC_t0_slow_fast}), converted to surface brightness. At late times, the black dots are above the blue line because they include the slow ejecta. The best fit $t_{0,\rm fast}$ is marked as a vertical dotted line, and this line intersects the black solid line at a level of $d\sigma/d\Omega(\theta)/\sigma_{\rm T}\approx3\times10^{-4}$.


\section{The slow ejecta}
\label{sec:slow}

\subsection{The slow ejecta particle size and scattering properties}

We detect a slow ejecta component composed of a tail and fan-like structure (see also \citealt{Li+2023Nature_DART_EjectaActiveAsteroid, Moreno+2023PSJ_DART_Impact_Ejecta_Observations_and_Models}). This component is responsible for the second plateau in the light curve, which started about 600--1000\,s after the impact and extends up to about 1\,day. From its brightness we get $S_{\rm ej, slow, B}\approx 0.0095$\,km$^{2}$. From the fact that the second plateau started declining at $t\sim1$\,day, and reached an effective surface area of $0.002$\,km$^{2}$ by day 8, we can estimate that the slowly-moving ejecta had typical projected velocities in the range of $\sim0.5$ to $\sim10$\,m\,s$^{-1}$, with roughly half the surface area with a projected velocity below $\sim4$\,m\,s$^{-1}$.
Since this velocity is estimated using the time it takes the flux to leave the photometric aperture, we estimate it may suffer from a factor of 2 uncertainty.

The time scale on which the ejecta become optically thin (i.e., $t_0$) is consistent with the fact that the maximum surface area in the larger photometric apertures is bigger than the one in the smaller apertures and converges for an aperture radius of about $10''$. Specifically, for $t_{0,{\rm slow}}\approx25$\,s (see Appendix~\ref{app:Model_t0}), the slow ejecta light curve is still not fully optically thin at 100\,s after the impact. Therefore, the flux in the larger aperture increases as the slow ejecta becomes optically thin, while the flux in the smaller aperture decreases as the fast ejecta leaves the aperture.

Assuming a projected velocity with a flat distribution between 0.5 and 10\,m\,s$^{-1}$, $\sqrt{\langle v_{\rm ej, slow}^{2}\rangle}\gtrsim 6$\,m\,s$^{-1}$, and from the kinetic energy conservation (Eq.~\ref{eq:EnergyCons}), we can put an upper limit on the mass of the slow ejecta $\lesssim 6\times10^{8}$\,kg (about 10\% of Dimorphos mass).
This limit is independent of the scattering properties of the ejecta.
Using this upper limit, along with the ratio between surface area and mass
we can put an upper limit on the average particle radius:
\begin{align}
    \tilde{r}_{\rm p, slow} \approx \frac{3 m_{\rm ej}}{4\rho S_{\rm eff}} \frac{d\sigma/d\Omega(\theta)}{\pi \tilde{r}_{\rm p}^{2}} \nonumber \\
    \lesssim 14 \Big(\frac{\rho}{3500\,{\rm kg\,m}^{-3}} \Big)^{-1} Q_{\rm scat}(\theta)\,{\rm m},
    \label{eq:rLimit}
\end{align}
where $Q_{\rm scat}(\theta)=d\sigma/d\Omega(\theta)/(\pi r_{\rm p}^{2})$, and $\theta=53.3$\,deg (i.e., the scattering cross section in units of the geometric cross-section), $r_{\rm p}$ is the scattering particle radius,
and $\rho$ is the particle density.
For the slow ejecta, we adopted $\rho=3500$\,kg\,m$^{-3}$ consistent with stony meteorites (\citealt{Britt+2003M&PS_StonyMeteroriteProperties}).
If the slow ejecta material is similar in scattering properties
to the asteroid's surface material, then its scattering efficiency,
as measured from the pre-impact light,
is about $0.008$, and hence $\tilde{r}_{\rm p, slow}\lesssim 0.11$\,m.
%
In HST images obtained months after the impact \cite{Jewitt+2023ApJ_DART_Impact_DimorphosBoulderSwarm} found dozens of slowly moving ($\sim0.3$\,m\,s$^{-1}$) point sources around Didymous.
Assuming they have the albedo of Didymos, their size is about 3\,m.
Since our upper limit on the particle size refers to the (reflected light-weighted) average particle size in the slow ejecta and not to the largest particles, our limit is consistent with their results.

Another limit on the particle size in the ejecta can be obtained from the smoothness of the ejecta.
Since the slow ejecta in the images look smooth
(i.e., we do~not see discrete particles) we get, for mono-size particles,
$r_{\rm p}\lesssim (S/[\pi N_{\rm p, slow}])^{1/2}$, where $N_{\rm p, slow}$ is the number of discrete particles in the slow ejecta.
For $N_{\rm p, slow}\gtrsim1000$, we get $r_{\rm p, slow}\lesssim 20$\,m.
However, the direct detection of meter-size particles by \cite{Jewitt+2010Natur_AsteroidCollision_P2010A2} supersedes this estimate.

By substituting Equation~\ref{eq:rLimit} and the Equation for the optical thickness time scale $t_{0}$ (Appendix~\ref{app:Model_t0}; Equation~\ref{eq:t0_mono}) and rearranging we get
\begin{eqnarray}
    \frac{d\sigma/d\Omega(\theta)}{\sigma_{\rm T}} \sim \frac{3}{2\pi} \frac{S_{\rm eff}}{v_{\rm ej}^{2} t_{0}^{2} (1-\cos\gamma) }
    \label{eq:epsilon}    
\end{eqnarray}
Here $\sigma_{\rm T}$ is the total cross-section for absorption and scattering (see Appendix~\ref{app:Model_t0}).
This equation provides an estimate of the scattering properties of the slow ejecta.
Given an upper limit on the slow-ejecta velocity, this can be used to set a lower limit on the scattering
efficiency of the mm-size (and above) particles in the slow ejecta (assuming $\gamma\sim1$):
\begin{equation}
    \frac{d\sigma/d\Omega(\theta)}{\sigma_{\rm T}}\gtrsim 0.07 \frac{S_{\rm eff}}{0.0095\,{\rm km}^{2}} \Big(\frac{v_{\rm ej}}{10\,{\rm m\,s}^{-1}} \frac{t_{0}}{24.5\,{\rm s}} \Big)^{-2}
\end{equation}
However, due to projection effects, the actual velocity may be somewhat larger than 10\,m\,s$^{-1}$.
Nevertheless, this suggests that the large particles in the slow ejecta
have scattering efficiency, which is at least a factor of a few larger compared to the pre-impact scattering efficiency of Didymos.
Interestingly Polishook et al. (2023) found that the spectral slope of the IR spectra of the slow ejecta component is bluer compared to the pre-impact spectra of Didymos.

\subsection{Tail and radiation pressure}
\label{sec:RadP}

About 24\,hr after the impact, a thin tail appeared
at a position angle of about 289\,deg (see also \citealt{Li+2023Nature_DART_EjectaActiveAsteroid, Moreno+2023PSJ_DART_Impact_Ejecta_Observations_and_Models}).
The position angle of the antisolar vector (at day 3 from impact) is 293\,deg,
and the position angle of the negative of the Heliocentric velocity vector is 224\,deg.
On day two, the tail extends over 200\,arcsec.
Assuming the tail is in the antisolar direction and given that the Sun-Didymos-Earth angle is 60\,deg,
this is translated to tail length of $l_{\rm tail}\approx1.2\times10^{7}$\,m.
The particles in the tail are presumably accelerated by solar radiation pressure. For spherical particles, this acceleration is 
given by
\begin{equation}
    a_{\rm R} \approx \frac{\pi r_{\rm p}^{2}}{4/3\pi \rho r_{\rm p}^{3}} \frac{L_{\odot}}{4\pi r^{2} c} = \frac{3 L_{\odot}}{16\pi\rho c r_{\rm p} r^{2}},
    \label{eq:RadPress}
\end{equation}
where $L_{\odot}$ is the Sun luminosity, and $c$ is the speed of light. The maximum acceleration is $\approx 2l_{\rm tail}/t^{2}\sim3\times10^{-4}$\,m\,s$^{-2}$.
Assuming the particles in the tail
have Regolith-like densities (\citealt{McKay+1972LPSC_LunarRegolithParticlesSizeDistribution}) of about 1,500\,kg\,m$^{-3}$, and
combining this with the measured maximum acceleration indicates
particles at the edge of the tail have a radius of about $7\,\mu$m,
with closer-to-Dimorphos parts of the tail containing bigger particles ($\sim0.1$\,mm).
This is consistent with the findings of \cite{Moreno+2023PSJ_DART_Impact_Ejecta_Observations_and_Models}.
We note that in such cases, infrared spectroscopy or photometry of the tail may reveal, via Mie scattering analysis (see Appendix~\ref{sec:Mie}), the size distribution and dielectric properties of the particles.
This is crucial for estimating the mass of the dust particles, as the actual scattering properties
of the particles may differ significantly from assumptions based on the global albedo of the asteroid (see, e.g., Figure~\ref{fig:Mie_Q_rp}).


We measure the surface brightness of the tail as a function of distance
from Didymos, and its integrated brightness, from $16''$ to $200''$ amounts
to $S_{\rm tail}\approx8\times10^{-3}$\,km$^{2}$.
Using Equation~\ref{eq:rLimit}, with the particle radius and surface brightness
as a function of distance from Didymos,
we estimate the mass of the tail as a function of distance.
The integrated mass of the tail amounts to $m_{\rm tail}\sim 800\,Q_{\rm scat, tail}^{-1}(\theta)$\,kg, with $\theta=53.3$\,deg,
independent of the particle density.
Assuming Mie scattering with $n+i\kappa=1.5+0.3i$ (appropriate for Silica dust; \citealt{Draine+1984ApJ_OpticalProperties_Graphite_Silicate_Grains}) and $\sim10\,\mu$m particles,
$Q_{\rm scat}(\theta)\sim10^{-3}$\,sr$^{-1}$ with weak dependency on the particle size (see Fig.~\ref{fig:Mie_Q_rp}),
the mass of the tail is $m_{\rm tail}\sim 8\times10^{5}$\,kg.

The fact that the tail is narrow suggests that the spread in the initial velocity of the particles in the tail is smaller than $0.3$\,m\,s$^{-1}$ (i.e., spreading by less than $1''$ over four days).
This velocity is comparable to the escape velocity from the Didymos system at the Dimorphos orbit.


Material ejected in such low velocities will become optically thin
(Equation~\ref{eq:t0_mono} with $\gamma=1$) at around 
\begin{equation}
    t_{0} \approx 10^{4} \Big(\frac{m_{\rm ej}}{8\times10^{5}\,{\rm kg}} \frac{\rho}{1.5\,{\rm g\,cm}^{-3}}\Big)^{1/2} \Big(\frac{\tilde{r}_{\rm p}}{20\,\mu{\rm m}} \Big)^{-1/2} \Big(\frac{v_{\rm ej}}{0.3\,{\rm m\,s}^{-1}} \Big)^{-1}\,{\rm s},
\end{equation}
where we normalize the density to the regolith density (\citealt{McKay+1972LPSC_LunarRegolithParticlesSizeDistribution}),
and we use an average particle radius of $20\,\mu$m.
Interestingly, at times between 1000\,s to about $10^{4}$\,s the light curve rises by (at least) $\sim0.001$\,km$^{2}$ ($\sim0.05$-magnitudes wiggle in Fig.~\ref{fig:DART_LAST_mag_logt_apersPhases}).
Although the observations were obtained at high airmass
(in the range of 2.2 to 3.8),
the fact that they are seen in
both the C28 and LAST data sets,
and the telescopes are separated by about 65\,km, which likely means that the variations are real.
%
Assuming that the wiggle in the light curve at $10^{4}$\,s corresponds to $t_{0}\sim3000$\,s and that the ejection velocity of the tail particles is between $0.1$ to $0.3$\,m\,s$^{-1}$, using Equation~\ref{eq:epsilon} we get $Q\sim10^{-3}$.
This is consistent with the expected scattering efficiency of $\mu$m-size Silica dust particles (\citealt{Draine+1984ApJ_OpticalProperties_Graphite_Silicate_Grains}; see Fig.~\ref{fig:Mie_Q_rp}).


The tail appears about 4\,deg off the antisolar direction towards the negative
of the heliocentric velocity vector.
This is presumably due to the Coriolis force that acts on the tail particles
in the non-inertial frame of the asteroid.
The Coriolis force acts in the direction of the negative of the Heliocentric velocity vector of
the asteroid, where its expected acceleration, in this direction, is $a_{\rm C} = 2\omega a_{\rm R} t$,
where $\omega\approx2\times10^{-7}$\,rad\,s$^{-1}$ is the angular velocity of the asteroid around the Sun,
and the expected (non-projected) angle of the tail from the antisolar direction towards the negative of the Heliocentric velocity vector is
\begin{equation}
    \tan{\xi} \approx \frac{a_{\rm C}t^{2} / 2}{a_{\rm R}t^{2} / 2} = \frac{\omega a_{\rm R} t^{3}}{a_{\rm R} t^{2} /2} = 2\omega t.
\end{equation}
This approximation is valid because the tail length is much smaller than the orbital size.
Since the tail is short we can neglect additional forces (e.g., Lorentz force and Poynting-Robertson drag).
Since the angle between the observer and target orbital plane, measured from the center
of the target was 49\,deg (at $t=3$\,day), then the expected tilt
is $\xi\sin{(49\,{\rm deg})}\cong4.8$\,deg, close to the observed angle of the tail.
At later times there is some evidence that the deviation is larger.
This may be related to the double tail reported in the {\it HST} images of the impact (e.g., \citealt{Li+2023Nature_DART_EjectaActiveAsteroid, Kim+Jewitt2023arXiv_DART_Impact_SIngleEjectionToExplainDoubleTail, Moreno+2023PSJ_DART_Impact_Ejecta_Observations_and_Models}).

There are two possible explanations for the double-tail appearance.
\cite{Kim+Jewitt2023arXiv_DART_Impact_SIngleEjectionToExplainDoubleTail} show that such a structure can be formed by the spread in the directions in which the tail's material is ejected.
An alternative explanation is due to an ejection of material at later times (a few days after the impact).
Such a model was suggested by \cite{Tancredi+2022arXiv_DART_Didymos_SeismicWavesDustLofting}.
Interestingly, the light curve of the DART impact (Fig.~\ref{fig:DART_LAST_mag_logt_apersPhases}) shows a brightening on day four after the impact, which can be consistent with such a late time ejection.

Next, the bulk of the resolved slow ejecta does not drift in the antisolar direction,
and, therefore, is not strongly affected by
the solar radiation pressure. This can be used to put a lower limit on the particle size (see also \citealt{Li+2023Nature_DART_EjectaActiveAsteroid, Moreno+2023PSJ_DART_Impact_Ejecta_Observations_and_Models}).
The images suggest that, around day 3--5, the curvature of the fans toward the antisolar direction
is smaller than $\mathcal{C}<1/10$.
Therefore, the particle acceleration due to the solar radiation pressure is:
$a_{\rm R, fans}\lesssim 2v/(\mathcal{C}t)$.
For $v=4$\,m\,s$^{-1}$ and $t=5$\,day we get $a_{\rm R, fan}\lesssim 2\times10^{-6}$\,m\,s$^{-2}$,
which in turn, using Equation~\ref{eq:RadPress}, gives $r_{\rm p, slow}\gtrsim 0.4$\,mm.
Substituting into Equation~\ref{eq:rLimit}, we can put a lower limit on the 
mass of the slow ejecta: $m_{\rm ej, slow}\gtrsim 1.8\times10^{5} Q_{\rm scat}^{-1}(\theta)$\,kg.
Assuming $Q_{\rm scat}\sim0.07$, the mass of the slow ejecta is $m_{\rm ej, slow}\gtrsim 3\times10^{6}$\,kg.
Furthermore, assuming the process responsible for the slow ejecta also created a crater whose depth-to-radius ratio is $d$, then the crater radius (from simple geometric considerations)
is
\begin{equation}
    r_{\rm crater}\sim \Big(\frac{m_{\rm ej}}{\pi \rho d}\Big)^{1/3},
\end{equation}
and with $d=1/3$, we get a lower limit of $r_{\rm crater}\gtrsim9$\,m.

\section{Conclusions}
\label{sec:conclusion}

We present UV, visible light, and near-IR ($z$-band) observations of the DART impact (\citealt{Rivkin+2021PSJ_DART_MissionRequirments}).
We use these observations to estimate the properties of the ejecta, which are summarized in Table~\ref{tab:sum}.
The observations suggest that the ejecta is composed of two main velocity components.
A fast component that contains gas
(\citealt{Shestakova+2023Icar_DART_Impact_Spectroscopy_EmissionLines}), and a slow component.
The two components seem to be well separated in velocity, and they seem to have different geometric appearances.
The slow ejecta contains both micro-meter size particles (that form the tail), as well as mm-sized (and larger, up to a few meters) particles that are only weakly affected by the solar radiation pressure.
Although some simulations predict that the highest velocities of the ejecta
will be of the order of 2\,km\,s$^{-1}$ (\citealt{Luther+2018M&PS_AsteroidsCollisions_DART_ImpactModeling}),
close to the observed value, as far as we know, these simulations
do~not predict an ejecta with bimodal velocity distribution.

An interesting question is which velocity component dominates
the momentum transfer gained by the impact.
Following \cite{Cheng+2023Nature_DART_MomentumTransfer}
we define the momentum gain by:
\begin{equation}
    \beta = 1 + \frac{M\Delta{\vec{v}} - m_{\rm imp}\vec{v}_{\rm imp} }{m_{\rm imp} (\widehat{E}\cdot\vec{v}_{\rm imp})\widehat{E}  } \approx 1 + \frac{P_{\rm ej}}{m_{\rm imp}v_{\rm imp}}.
  \label{eq:beta}
\end{equation}
Here, $M\Delta\vec{v}$ is the momentum change of Dimorphos due to the impact,
$\widehat{E}$ is the net ejecta momentum direction,
and $P_{\rm ej}$ is the ejecta momentum.
The left equality is the $\beta$ definition used by the DART team (\citealt{Cheng+2023Nature_DART_MomentumTransfer}),
while the right-hand side approximation is written in terms of the ejecta mass and velocity, which are measured in this work.
Given the observed momentum gain ($\beta\sim4$; \citealt{Cheng+2023Nature_DART_MomentumTransfer}), and
the approximate velocity of the slow ejecta ($\sim4$\,m\,s$^{-1}$), a slow-ejecta mass of the order of
$\sim 3\times10^{6} (v_{\rm ej,slow}/4\,{\rm m\,s}^{-1})^{-1}$\,kg is needed
in order to explain the observed value of $\beta\sim4$.
This is consistent with the limits we find.
We note that \cite{Kim+Jewitt2023arXiv_DART_Impact_SIngleEjectionToExplainDoubleTail} estimated the total mass of the slow ejecta to be $\sim2\times10^{7}$\,kg (Didymos albedo was assumed).
This estimate suggests that the slow ejecta dominates the momentum gain of the impact.
On the other hand, if the fast-ejecta dominated the
momentum-transfer gain, then a fast-ejecta mass of
the order of $7\times10^{3}$\,kg is required.
A possible way to solve the question of which ejecta component dominated the momentum-transfer gain, is to better estimate the fast-ejecta mass.
This may be possible by a non-steady state modeling of the radiative processes of the fast-ejecta gas.


%
\begin{table*}
\centering
\begin{tabular}{ll}
\hline
Property  & Limits \\
\hline\hline
$v_{\rm ej, fast}$ & $\approx1600\pm200$\,m\,s$^{-1}$ [M]; with tail down to 500\,m\,s$^{-1}$ [LC] \\
$m_{\rm ej, fast}$ & $\lesssim1\times10^{4}$\,kg [KE]\\
$r_{\rm p, fast}$  & $\lesssim0.2\,\mu$m [SP]; $\lesssim$ few~$\mu$m [SM] \\
$\kappa_{\rm visible}$           & $\gtrsim10^{4}$ [SP] \\
$\kappa_{\rm UV}$                & $\gtrsim10^{4}$ [SP] \\
\hline
$v_{\rm ej, slow}$ & $\sim0.1$--$10$\,m\,s$^{-1}$ [LC+M]\\
$m_{\rm ej, slow}$ & $\gtrsim2\times10^{5}\,Q_{\rm scat, slow}^{-1}$\,kg [A+SM]; $\gtrsim9\times10^{5}$\,kg (with $Q$) \\
$r_{\rm p, slow}$  & $\gtrsim1$\,mm [A]; $\lesssim14$\,m [IS]; $\lesssim3$\,m [IS] (with $Q$) \\
$Q_{\rm scat,slow}(\theta)$ & $\gtrsim 0.07$ [SM, SP, LC]\\
\hline
$v_{\rm ej, tail}$ & $\lesssim0.1$\,m\,s$^{-1}$ [A] \\
$m_{\rm ej, tail}$ & $\sim800 Q_{\rm scat, tail}^{-1}(\theta)$\,kg [A] \\
$Q_{\rm scat, tail}$ & $\sim10^{-3}$ [SP]\\
$r_{\rm p, tail}$  & $\gtrsim7\,\mu$m [A]; $\lesssim1000\,\mu$m [A] \\
\hline
\end{tabular}
\caption{\label{tab:sum} 
Summary of estimated ejecta components and properties.
For each limit, we also provide the method by which it was obtained:
LC - light curve; KE - energy conservation; M - Motion in images; SM - surface area to mass; SP - scattering properties; T - Optical thickness; A - Acceleration due to radiation pressure; IS - Smoothness in images. See the text for details.
}
\end{table*}


%
\section*{Acknowledgements}

We are thankful to the staff of the Tel-Aviv University Wise Observatory; Didymos is the first target simultaneously observed by the LAST array at the Weizmann Astrophysical Observatory and telescopes at the Wise Observatory.
E.O.O. is grateful for the support of
grants from the 
Benoziyo Center,
Willner Family Leadership Institute,
Ilan Gluzman (Secaucus NJ), Madame Olga Klein - Astrachan,
Minerva Foundation,
Israel Science Foundation,
BSF-NSF, Israel Ministry of Science, Weizmann-MIT,
and the Rosa and Emilio Segr\'e Research Award.
DP is thankful to the DART mission investigation team for fruitful collaboration.
N.L.S. is funded by the Deutsche Forschungsgemeinschaft (DFG, German Research Foundation) via the Walter Benjamin program – 461903330.

\section*{Data Availability}

The data presented in this paper is available in the electronic tables, while the code is accessible via GitHub\footnote{\url{https://github.com/EranOfek/AstroPack}}.

\onecolumn

\appendix

\section{Derivation of the optical-thickness time scale}
\label{app:Model_t0}

Here, we develop a simple optical-thickness model, taking into account some of the geometrical properties of the DART impact.
The observed flux in some band $\rm{x}$ is given by
\begin{equation}
    f_{\rm x}=f_{\odot,\rm{x}}\Delta\Omega_p\mathbb{P},
\end{equation}
where $f_{\odot,\rm{x}}$ is the flux from the Sun reaching the ejecta, $\Delta\Omega_p$ is the solid angle occupied by the ejecta, as seen by the observer, and $\mathbb{P}$ is the probability that a photon that reaches the ejecta will be scattered into this solid angle. It is convenient to define the effective trajectory length of a photon within the ejecta, $l_{\rm eff}$ though
\begin{equation}
    \mathbb{P}\equiv\frac{1}{\sigma_{\rm T}}\frac{d\sigma}{d\Omega}\frac{l_{\rm eff}}{l},
\end{equation}
where $\sigma_{\rm T}$ is the total (scattering plus absorption) cross-section, and $l$ is the photon mean free path. In the limit that the ejecta is optically thick, we have $l_{\rm eff}\approx l$, and in the limit that the ejecta is optically thin, we have $l_{\rm eff}\approx L$, where $L$ is the size of the ejecta.
Before we derive a more realistic model, we would like to give some intuition regarding the asymptotic behavior of the flux,
and this will be done by assuming some simple box-like geometry (that will be relaxed later on).
In this case, we can write
\begin{eqnarray}
    \frac{f_{\rm x}}{f_{\odot,\rm{x}}}=\frac{1}{\sigma_{\rm T}}\frac{d\sigma}{d\Omega}\frac{l_{\rm eff}}{l}\Delta\Omega_p\\
    =\begin{cases}
\frac{1}{\sigma_{\rm T}}\frac{d\sigma}{d\Omega}\Delta\Omega_p
=\frac{1}{\sigma_{\rm T}}\frac{d\sigma}{d\Omega}\frac{A_p}{\Delta^2}\propto t^2 & t\rightarrow0, \\
nL\frac{d\sigma}{d\Omega}\Delta\Omega_p =N_p\frac{d\sigma}{d\Omega}\frac{1}{\Delta^2}=\rm{const}.& t\rightarrow\infty,
\label{eq:fxfxsun}
\end{cases}
\end{eqnarray}
where $\Delta$ is the observer-ejecta distance,
and we used $\Delta\Omega_p=A_p/\Delta^2$, $A_p=L^2$ is the geometric area of the ejecta, $N_p=nL^3$ is the number of particles in the ejecta,
and $n$ is the particle density in the ejecta.
Here, the $t\rightarrow\infty$ case is provided just to show that the flux goes into a constant value. Below, we are using a more realistic geometrical model.

To estimate $l_{\rm eff}$, we require a geometrical model for the ejecta. The fast ejecta of the DART impact looks like a spherical section, with a half-opening angle $\gamma\approx1$, width/radius $\delta\approx0.5$ and a radial velocity $v_{\rm ej}$. The maximal cylindrical radius of the shell is $w_{2}=  v_{\rm ej}t\sin(\gamma)$ and we define the thickness of the spherical shell by $w_{1}=\delta v_{\rm ej}t$. Since the DART impact vector was roughly at $\alpha_{\rm DAO}\approx90$\,deg from our line of sight and the Sun-Didymos-Earth angle $\theta=53.3\,\rm{deg}$, we use the approximation in which the Sun shines in parallel with the section axis, and photons have to get out, to Earth, of the vertical direction. This approximation holds for $\delta\gtrsim|\alpha_{\rm DAO}-\pi/2|\approx0.1$ (here $\alpha_{\rm DAO}$ and $\delta$ are in radians).
Our approximation requires that
$\sigma_{\rm abs}\gg\sigma_{\rm scat}$, which is consistent with the observations, as our analysis shows. 
This requirement means that each photon is scattered not more than one time.


%
Next, we define a cylindrical coordinate system in which we do the calculations,
where $z$ is the cylinder axis, $r$ is the cylinder radius coordinate, and $\phi$ is the cylinder angle coordinate. 
We can write 
\begin{equation}
    l_{\rm eff}=\frac{1}{\pi w_2^2}\int_{0}^{2\pi}\int_{0}^{w_2}\int_{z_1(r)}^{z_2(r)}e^{-\frac{z-z_1(r)}{l}}e^{-\frac{d(z,r,\phi)}{l}}rdr\,dz \,d\phi,
\end{equation}
where the first term of the integrand is the probability to scatter within the ejecta (as the photon is moving along the $z$ direction and the ejecta spans the region $[z_1(r),z_2(r)]$), and the second term of the integrand is the probability that the scattered photon will not suffer more scattering within the ejecta, where $d(z,r,\phi)$ is the distance the scattered photon has to travel within the ejecta. It is more convenient to use the outer radius of the shell $R_2=w_2/\sin(\gamma)$ and the inner radius of the shell $R_1=R_2-w_1$ for the calculation. We obtain 

\begin{eqnarray}
    z_1(r)=R_2\left(1-\sqrt{1-\left(\frac{r}{R_2}\right)^2}\right),\\
    z_2(r)-z_1(r)= \begin{cases}
R_2\left(\sqrt{1-\left(\frac{r}{R_2}\right)^2}-\frac{r}{R_2}\cot(\gamma)\right)-R_1\left(\sqrt{1-\left(\frac{r}{R_1}\right)^2}-\frac{r}{R_1}\cot(\gamma)\right) & r\le\Delta r, \\
R_2\left(\sqrt{1-\left(\frac{r}{R_2}\right)^2}-\frac{r}{R_2}\cot(\gamma)\right) & r>\Delta r, \\
\end{cases}
\end{eqnarray}
where $\Delta r=R_1\sin(\gamma)$ is the maximal cylindrical radius of the inner shell. To calculate $d$, we assume that the scattering event was at $(z,r,\phi)$ and assume Earth is along the $x$-axis (perpendicular to the $z$ axis). Then, the distance to Earth within the ejecta is given by
\begin{eqnarray}
    d(z,r,\phi)= \begin{cases}
\sqrt{\tilde{r}^2-r^2\cos^2(\phi)}-r\sin(\phi) & z\le w_1, \\
\sqrt{\tilde{r}^2-r^2\cos^2(\phi)}-r\sin(\phi)-2R_2\sqrt{\left(\frac{R_1}{R_2}\right)^2-\left(1-\frac{z}{R_2}\right)^2} & z>w_1, \\
\end{cases}
\end{eqnarray}
where 
\begin{eqnarray}
    \tilde{r}(z)=
    \begin{cases}
        R_2\sqrt{1-\left(1-\frac{z}{R_2}\right)^2} & z\le R_2 (1-\cos(\gamma)), \\
        R_2\left(1-\frac{z}{R_2}\right)\tan(\gamma) & z>R_2 (1-\cos(\gamma)),
    \end{cases}
\end{eqnarray}
is the maximal cylindrical radius at $z$. 

The volume of the shell is given by
\begin{equation}
    V = \frac{2}{3}\pi v_{\rm ej}^{3}t^{3} (1-\cos{\gamma}) (1 - (1-\delta)^{3}).
\end{equation}
For mono-size particles, the number density is:
\begin{equation}
    n=\frac{9}{8\pi^2}\frac{M_{\rm ej}}{\rho r_{\rm p}^{3} v_{\rm ej}^{3}t^3[1-\cos(\gamma)][1-(1-\delta)^3]},
\end{equation}
allowing us to write 
\begin{eqnarray}
    \frac{R_2}{l}=\left(\frac{t_0}{t}\right)^2\frac{1}{1-(1-\delta)^3},\\
    \frac{R_1}{l}=\left(\frac{t_0}{t}\right)^2\frac{1-\delta}{1-(1-\delta)^3},
\end{eqnarray}
with
\begin{equation}
    t_{0} = \sqrt{\frac{9 m_{\rm ej} \sigma_{\rm T} }{8\pi^2 \rho r_{\rm p}^3  v_{\rm ej}^{2} (1-\cos(\gamma))} }.
    \label{eq:t0_mono}
\end{equation}

\begin{figure}
\centering
\includegraphics[width=\linewidth]{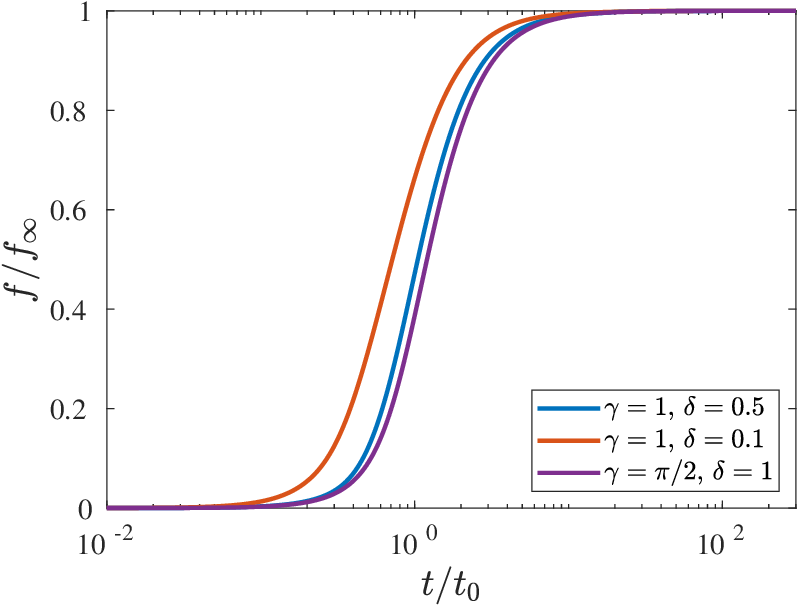}
\caption{The theoretical rising light curve due to optical thickness model for selected values of $\gamma$ and $\delta$.}
\label{fig:L_t_sphere}
\end{figure}

The value of $l_{\rm eff}$ can be calculated numerically as a function of $t/t_0$ (for a choice of $\gamma$ and $\delta$), allowing us to plot $f_{\rm x}/f_{\rm{x},\infty}$, where $f_{\rm x,\infty}$ is the observed flux at $t\rightarrow\infty$, in Figure~\ref{fig:L_t_sphere}. 
%

For resonance scattering, this model assumes that the absorption cross-section for light entering the ejecta,
and going out is roughly the same.

\subsection{Equations for particles with size distribution}
\label{sec:ModelSizeDist}

So far we have written the equations for mono-size particles. For particles with differential size distribution $dn/dr$, we can use the following equations:
\begin{equation}
    n = \frac{1}{V} \frac{M}{\frac{4}{3} \pi \rho \int{\frac{dn}{dr}r^{3}dr}} \int{\frac{dn}{dr}dr},
\end{equation}
and in this case, the effective absorption cross-section is
\begin{equation}
    \sigma_{\rm abs}=\int{\frac{dn}{dr}\sigma_{\rm abs}(r) dr},
\end{equation}
while $t_{0}$ is given by:
\begin{equation}
    t_{0}^{2} = \frac{9 m_{\rm ej}}{8\pi^{2}\rho v_{\rm ej}^{2} (1-\cos\gamma)\int{\frac{dn}{dr}r^{3}dr } } \int{\frac{dn}{dr} dr }\int{\frac{dn}{dr}\sigma_{\rm abs}(r)dr }.
\end{equation}

\subsection{The model surface brightness}
\label{sec:SurfB}
We next calculate the surface brightness for this geometrical model. We need to consider the range $0\le z\le R_2-R_1\cos(\gamma)$ and $0\le h \le \tilde{r}(z)$, where $z$ is the distance from the edge of the ejecta closest to the Sun and $h$ is the distance from the section axis. At early times, when the ejecta is not resolved, we can only measure the mean surface brightness by 
\begin{equation}
    I_{\rm x}=\frac{f_{\rm x}}{\Delta \Omega_p}=\Delta^2\frac{f_{\rm x}}{A_p},
\end{equation}
where $A_p$ is given in our model by
\begin{equation}
    A_p=2\int_{0}^{R_2-R_1\cos(\gamma)}\tilde{r}(z)dz\equiv R_2^2f_S(\gamma,\delta)=v_{\rm ej}^2t^2f_S(\gamma,\delta),
\end{equation}
where $f_S$ is some numerical factor ($\approx0.89$ for $\gamma=1$ and $\delta=0.5$). At late times, when the ejecta is resolved, we may measure the surface brightness as a function of angular location within the ejecta, given by:
\begin{eqnarray}
    I_{\rm x}(h,z)=A
    \begin{cases}
        \int_{0}^{x_1}e^{-\frac{z-z_1(r(x,h))}{l}}e^{-x/l}dx+\int_{x_2}^{2\sqrt{\tilde{r}(z)^2-h^2}}e^{-\frac{z-z_1(r(x,h))}{l}}e^{-\frac{x-x_2+x_1}{l}}dx & z> w_1\;\textrm{and}\;h<\tilde{r}_1, \\
        \int_{0}^{2\sqrt{\tilde{r}(z)^2-h^2}}e^{-\frac{z-z_1(r(x,h))}{l}}e^{-\frac{x}{l}}dx & \rm{otherwise},
    \end{cases}
    \label{eq:S}
\end{eqnarray}
where $x$ is the distance along the ejecta as seen from Earth, related to $r$ through
\begin{equation}
    r=\sqrt{\tilde{r}^2+x^2-2x\sqrt{\tilde{r}^2-h^2}},
\end{equation}
$x_1$ and $x_2$ are the coordinates of the inner shell, given by
\begin{eqnarray}
    x_1=\sqrt{\tilde{r}^2-h^2}-\sqrt{\tilde{r}_1^2-h^2}, \\
        x_2= \sqrt{\tilde{r}^2-h^2}+\sqrt{\tilde{r}_1^2-h^2},
\end{eqnarray}
and $\tilde{r}_1$ is the maximal cylindrical radius of the inner shell:
\begin{eqnarray}
    \tilde{r}_1(z)=R_2\sqrt{\left(\frac{R_1}{R_2}\right)^2-\left(1-\frac{z}{R_2}\right)^2}.
\end{eqnarray}
To determine the constant $A$ we note that for $t\rightarrow0$ the surface brightness approaches the constant value, $f_{\odot,\rm x}\sigma_{\rm T}^{-1}d\sigma/d\Omega$, obtained at $h=z=0$. From Equation~\eqref{eq:S} we find for $t\rightarrow0$ that $I_{\rm x}(0,0)\rightarrow\int_{0}^{2\tilde{r}}e^{-x/l}dx\approx l$, such that 
\begin{eqnarray}
    A=f_{\odot,\rm x}\frac{1}{l\sigma_{\rm T}}\frac{d\sigma}{d\Omega}.
\end{eqnarray}
The maximal surface brightness at any time is obtained at $h=0$. 
The surface brightness at $h=0$ is given in the limit $t\rightarrow\infty$, as
\begin{eqnarray}
    I_{\rm x}(0,z)=2f_{\odot,\rm x}\frac{1}{l\sigma_{\rm T}}\frac{d\sigma}{d\Omega}
    \begin{cases}
        \tilde{r}-\tilde{r}_1 & z> w_1, \\
        \tilde{r} & z\le w_1,
    \end{cases}
\end{eqnarray}
and the maximum of this function is given by
\begin{eqnarray}
    I_{\rm x,\max}=2f_{\odot,\rm x}\frac{1}{\sigma_{\rm T}}\frac{d\sigma}{d\Omega}\frac{R_2}{l}
    \begin{cases}
        \sqrt{1-\left(1-\delta\right)^2} & 1-\delta>\cos(\gamma), \\
        \sin(\gamma) & 1-\delta\le \cos(\gamma),
    \end{cases}
\end{eqnarray}
which can be written as
\begin{eqnarray}
    I_{\rm x,\max}=2 f_{\odot,\rm x}\frac{1}{\sigma_{\rm T}}\frac{d\sigma}{d\Omega} \left(\frac{t_0}{t}\right)^2\frac{1}{1-\left(1-\delta\right)^3}
    \begin{cases}
        \sqrt{1-\left(1-\delta\right)^2} & 1-\delta>\cos(\gamma), \\
        \sin(\gamma) & 1-\delta\le \cos(\gamma).
    \end{cases}
\end{eqnarray}

The ratio between $I_{\rm x, max}$ and $I_{\rm x}$ in the optically thin limit is
given by:
\begin{eqnarray}
    \frac{I_{\rm x, max}}{I_{\rm x}} = \frac{2}{1-(1-\delta)^{3}} 
    \begin{cases}
        \sqrt{1-\left(1-\delta\right)^2} & 1-\delta>\cos(\gamma), \\
        \sin(\gamma) & 1-\delta\le \cos(\gamma).
    \end{cases}
\end{eqnarray}
For $\gamma=1$ and $\delta=0.5$, this ratio equal to $1.92$.

\begin{figure}
\centering
\includegraphics[width=\linewidth]{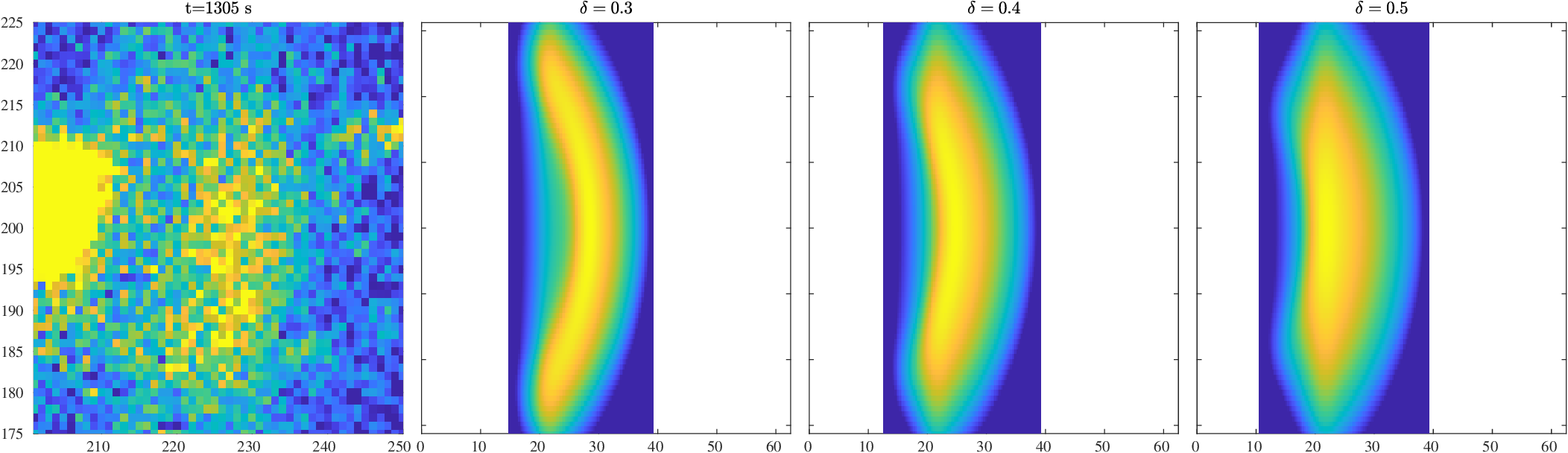}
\caption{Didymos coadded image taken at $t=1305$\,s compared with the expected 2D surface brightness of a uniform density spherical cup model with $\gamma=1$ and $\delta$ of 0.3, 0.4, and 0.5.}
\label{fig:SurfaceBrightnessModel2D}
\end{figure}

Figure~\ref{fig:SurfaceBrightnessModel2D} shows the fast ejecta coadded image taken at $t=1305$\,s compared with our surface brightness model for a uniform density spherical cup model with $\gamma=1$ and $\delta$ of 0.3, 0.4, and 0.5.
$\gamma=1$ and $\delta=0.5$ provide a reasonable match to the data.

\section{Mie Scattering interpretation}
\label{sec:Mie}

Prior to the realization that the light from the fast ejecta is due to emission lines from neutral gas, we attempted an interpretation based on the assumption that the fast ejecta is composed of scattering particles only.
This analysis found out that the particles must be small (below a few microns) and that Mie-scattering is the relevant scattering process.
The observations of \citealt{Shestakova+2023Icar_DART_Impact_Spectroscopy_EmissionLines} clearly show that the light of the fast ejecta is from resonance scattering of neutral atoms, and therefore, Mie scattering is not relevant (at least not for the majority of the observed light from the fast ejecta).
Here we give a brief version of this analysis because we believe it is useful due to the following reasons:
(i) This analysis shows, irrespective of the \cite{Shestakova+2023Icar_DART_Impact_Spectroscopy_EmissionLines} observations, that the particles in the fast ejecta must be smaller than $0.2\,\mu$m,
and that the mass of the fast ejecta must be dominated by particles smaller than $0.02\,\mu$m.
(ii) This line of analysis may be useful in the future for analyzing infrared spectra of tails.

The mass of the fast ejecta, $m_{\rm ej, fast}$, can be constrained from energy conservation, $m_{\rm ej, fast}v_{\rm ej, fast}^{2}\lesssim m_{\rm imp}v_{\rm imp}^{2}$, leading to $m_{\rm ej, fast}\lesssim10^{4}$\,kg.
Here the impactor mass is $m_{\rm imp}\cong579.4$\,kg, and impact velocity $v_{\rm imp}\cong 6145$\,m\,s$^{-1}$ (\citealt{Cheng+2023Nature_DART_MomentumTransfer}).
The upper limit on the mass, along with the measured effective surface area
of the ejecta gives us an upper limit on the typical particle radius,
$r_{\rm p}\lesssim 3 m_{\rm ej}/(4\rho S_{\rm eff})\sim 400\,\mu$m
(for a 100\% geometric scattering efficiency).
For the fast ejecta we assume $\rho=1500$\,kg\,m$^{-3}$ (Similar to regolith \citealt{McKay+1972LPSC_LunarRegolithParticlesSizeDistribution}).

For clarity, we first use a highly simplified description of the expanding fast ejecta: A spherical cloud expanding at $v_{\rm ej, fast}$, composed of spherical particles with a dielectric constant of $n+i\kappa \approx 1.5+0.3i$ (i.e., silica dust; \citealt{Draine+1984ApJ_OpticalProperties_Graphite_Silicate_Grains}) and two discrete radii.
This is because, as we show below, the scattering and absorption of light must be dominated by particles of different sizes. Later on, we present a more elaborate model, where the geometry is of an expanding shell consistent with the resolved images, the particle size distribution is continuous, and the dielectric constant is considered a free parameter. The key results obtained using this model and analysis are similar to those obtained using the simplified analysis.

\begin{figure}
\centering
\includegraphics[width=\linewidth]{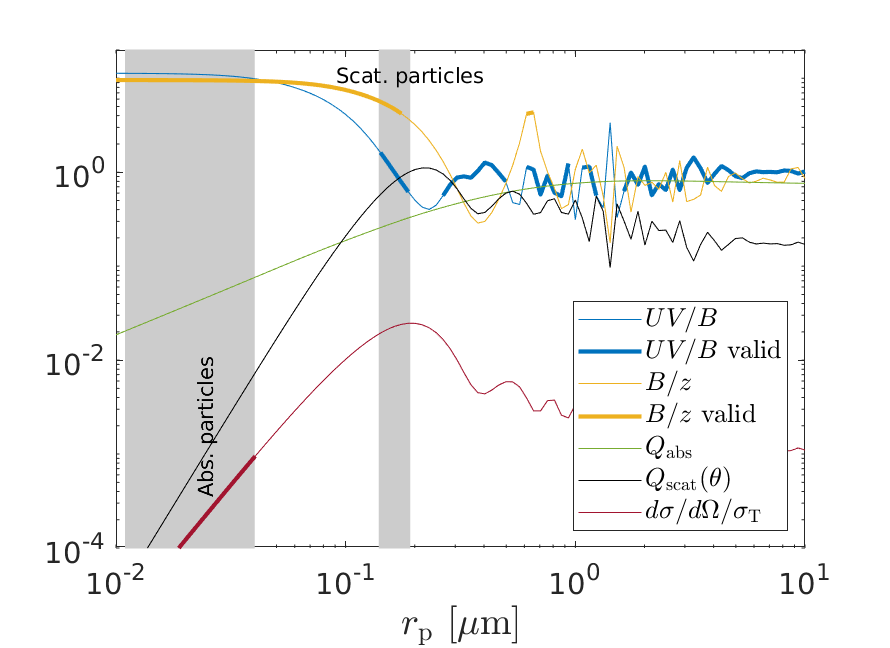}
\caption{The Mie scattering properties as a function of particle radius for $n+i\kappa = 1.5+i0.3$.
The blue line represents the calculated $UVW2$ to $B_{\rm p}$ band ratio,
while the yellow line is the same but for the $B_{\rm p}$ vs. $z$ ratio.
This calculation takes into account the spectral transmission as well as the solar spectrum in the band.
The heavy lines (and gray patches) show the regions that satisfy the observational constraints.
The green line shows
the absorption efficiency
($Q_{\rm abs}$; absorption cross section divided by $\pi r_{\rm p}^{2}$),
while the black line is for the 
scattering efficiency ($Q_{\rm scat}(\theta)=d\sigma/d\Omega(\theta)/(\pi r_{\rm p}^{2})$),
multiplied by 100.
The red line shows $d\sigma/d\Omega(\theta)/\sigma_{\rm T}$.}
\label{fig:Mie_Q_rp}
\end{figure}

The similarity of the maximal effective areas in the blue and UV bands, $S_{\rm eff, fast, B}\approx 0.0125$\,km$^{2}$ and $S_{\rm eff, fast, UVW2}\sim0.014$\,km$^{2}$, and much smaller effective area at near-infrared, $S_{\rm eff, fast, z}\lesssim0.002$\,km$^{2}$, implies that the size of the particles dominating the scattering is similar to the $B$-band radiation wavelength -- for particles with much larger radii the effective area would be similar at all wavelengths, while for particles with much smaller radii, the effective area in the blue and UV bands would be significantly different. 
This is demonstrated in
Figure~\ref{fig:Mie_Q_rp} that shows that the radius of particles dominating the Mie scattering is $r_{p,\rm scat}\approx0.1\,\mu$m. 

When the cloud is optically thin the effective area of the scattering particles is $N_{\rm scat} d\sigma/d\Omega(\theta)$, where $N_{\rm scat}$ is the number of scattering particles and $d\sigma/d\Omega(\theta)$ is their differential cross section for scattering in our direction ($\theta=53.3$\,deg). Using $d\sigma/d\Omega\approx0.02(\pi r_p^2)$ (see Figure~\ref{fig:Mie_Q_rp}) we find $N_{\rm scat}\sim2\times10^{19}$, and 
a total mass of the scattering particles of $M_{\rm scat}\sim 150$\,kg.

Had the debris cloud been composed only of the $r_{p,\rm scat}\approx0.1\,\mu$m {\it scattering} particles dominating the scattering, it would have become optically thin when $n_{\rm scat}\sigma_{\rm abs,scat}L\approx1$, where the ejecta length scale is $L\approx v_{\rm ej, fast}t_{0,\rm scat}$, the particle number density is $n_{\rm scat}\approx N_{\rm scat}/(4\pi L^3/3)$, $t_{0,\rm scat}$ is the $t_{0}$ of the scattering particles, and $\sigma_{\rm abs,scat}=\pi r_{p,\rm scat}^2$ (the absorption cross section is similar to the geometric cross section for $2\pi r_p\gtrsim \lambda$; see Figure~\ref{fig:Mie_Q_rp}), yielding $t_{0,\rm scat}\approx 0.2$\,s (see Fig.~\ref{fig:EjectaSurfaceBrightness}). The observed value of $t_{0,\rm fast}$, which is an order of magnitude larger than $t_{0,\rm scat}$, implies that the effective absorption cross-section per scattering particle should be two orders of magnitude larger than the scattering particles' geometric cross-section, $\sigma_{\rm abs,eff.}\approx 10^2 \pi r_{p,\rm scat}^2$. This implies that the cloud contains a mass $M_{\rm abs}\sim 10^2 M_{\rm scat}$ of particles of size $r_{p,\rm abs}\ll r_{p,\rm scat}$, that dominates the absorption and do not contribute significantly to the scattering.

Small particles, $r_{p,\rm abs}\ll r_{p,\rm scat}\approx\lambda$, may dominate the absorption while contributing little to the scattering since their absorption cross section is suppressed compared to their geometric cross section by a factor of $r_{p,\rm abs}/\lambda$ (Figure~\ref{fig:Mie_Q_rp}), while their scattering cross section is suppressed by a factor of $(r_{p,\rm abs}/\lambda)^2$. The total contribution of such particles to the absorption is independent of their size and depends only on their total mass, since the number of particles is $\propto M_{\rm abs}/r_{p,\rm abs}^3$ while their absorption cross section is $\approx \pi r_{p,\rm abs}^2(r_{p,\rm abs}/\lambda)\propto r_{p,\rm abs}^3$. The observed $\sigma_{\rm abs,eff.}\approx 10^2 \pi r_{p,\rm scat}^2$ therefore implies $M_{\rm abs}\sim 10^2 M_{\rm scat}\sim 1.5\times10^4$\,kg,
which is of the order of magnitude of the upper limit on the fast ejecta mass.
Assuming a factor of about 3 uncertainty on the measurement of $d\sigma/d\Omega/\sigma_{\rm T}$ this is translated to about $m_{\rm ej, fast}\gtrsim5000$\,kg.
In this case, the mass of the fast ejecta is close to the maximum allowed by energy conservation, $\approx 10^4$\,kg, and most of this mass is composed of small particles, with $r_{p,\rm abs}\ll0.1\,\mu$m, and that $\sim 10^2$\,kg of the mass is carried by particles with $r_{p,\rm scat}\approx 0.1\,\mu$m that dominate the scattering.

We fitted a simple Mie-scattering model\footnote{\url{https://www.mathworks.com/matlabcentral/fileexchange/36831-matscat?requestedDomain=}}
to the observational constraints.
Our model includes spherical particles
with refractive index $n+i\kappa$ which is wavelength-independent,
reflecting Sun light at angle $\theta=53.3$\,deg.
The particles follow a differential size distribution
of the form
\begin{equation}
    \frac{dN}{dr} \propto r^{-\alpha},
\end{equation}
where $\alpha$ is the power law index of the size distribution,
between minimum and maximum radii $R_{\rm min}$ and $R_{\rm max}$, respectively.
We use our geometrical model (\S\ref{sec:ModelSizeDist})
with $\gamma=1$ and $\delta=0.5$.
The observational constraints include:\\
(1) $m_{\rm ej, fast}<10^{4}$\,kg;\\
(2) $S_{\rm eff, fast, B}=0.0125$\,km$^{2}$;\\
(3) $0.33<Q_{\rm scat, UVW2}/Q_{\rm scat, B}<3$;\\
(4) $Q_{\rm scat, B}/Q_{\rm scat, z}>4$;\\
(5) $8.5\times10^{-5}<d\sigma/\Omega(\theta)/\sigma_{\rm T}<1.05\times10^{-3}$\,sr$^{-1}$.   

Our model assumes $\theta=53.3$\,deg
and $\sigma_{\rm T}\gg \sigma_{\rm scat}$,
where $\sigma_{\rm scat}$ is the total scattering cross-section
of the particle.
Given all these assumptions and additional uncertainty in the value of the observed parameters, we adopted a wide range for the observed parameters.
In all cases, we assumed $\rho=1500$\,kg\,m$^{-3}$ (\citealt {McKay+1972LPSC_LunarRegolithParticlesSizeDistribution}).
We scanned over the following parameters:
(i) $n$ in the range of 1.05 to 2.5 in steps of 0.025.
(ii) $\kappa$ in the range of 0.01 to 1 with 25 logarithmic steps.
(iii) $R_{\rm min}$ in the range of $0.01$ to $0.1\,\mu$m
in 12 logarithmic steps.
(iv) $R_{\rm max}$ in the range of $0.1$ to $0.6\,\mu$m in 20
logarithmic steps,
and (v) $\alpha$ in the range of $1.6$ to $6$ with step of $0.2$.

This model constrains some of the parameters very well.
For example, larger $R_{\rm max}$ will make the $z$ band reflectance higher
and will violate the observations.
This requires $R_{\rm max}\approx0.2\,\mu$m.
$\alpha\approx4.6$ is required in order to satisfy the observations.
Most importantly, based on this simple model, the mass of the fast
ejecta needs to be close to the upper limit on the fast ejecta mass
(from energy conservation), and we find $m_{\rm ej, fast}\sim9000$\,kg.
Given the \cite{Shestakova+2023Icar_DART_Impact_Spectroscopy_EmissionLines} observations, this estimate is not valid.
However, the limits on the particle sizes are still valid,
and they are independent of the emission lines observation.

\bibliographystyle{mnras}
\bibliography{papers.bib}

\begin{thebibliography}{}
\makeatletter
\relax
\def\mn@urlcharsother{\let\do\@makeother \do\$\do\&\do\#\do\^\do\_\do\%\do\~}
\def\mn@doi{\begingroup\mn@urlcharsother \@ifnextchar [ {\mn@doi@}
  {\mn@doi@[]}}
\def\mn@doi@[#1]#2{\def\@tempa{#1}\ifx\@tempa\@empty \href
  {http://dx.doi.org/#2} {doi:#2}\else \href {http://dx.doi.org/#2} {#1}\fi
  \endgroup}
\def\mn@eprint#1#2{\mn@eprint@#1:#2::\@nil}
\def\mn@eprint@arXiv#1{\href {http://arxiv.org/abs/#1} {{\tt arXiv:#1}}}
\def\mn@eprint@dblp#1{\href {http://dblp.uni-trier.de/rec/bibtex/#1.xml}
  {dblp:#1}}
\def\mn@eprint@#1:#2:#3:#4\@nil{\def\@tempa {#1}\def\@tempb {#2}\def\@tempc
  {#3}\ifx \@tempc \@empty \let \@tempc \@tempb \let \@tempb \@tempa \fi \ifx
  \@tempb \@empty \def\@tempb {arXiv}\fi \@ifundefined
  {mn@eprint@\@tempb}{\@tempb:\@tempc}{\expandafter \expandafter \csname
  mn@eprint@\@tempb\endcsname \expandafter{\@tempc}}}

\bibitem[\protect\citeauthoryear{{A'Hearn} et~al.,}{{A'Hearn}
  et~al.}{2005}]{AHearn+2005Sci_DeepImpact_Tempel1_Comet_ObservationsReview}
{A'Hearn} M.~F.,  et~al., 2005, \mn@doi [Science] {10.1126/science.1118923},
  \href {https://ui.adsabs.harvard.edu/abs/2005Sci...310..258A} {310, 258}

\bibitem[\protect\citeauthoryear{{Ben-Ami} et~al.,}{{Ben-Ami}
  et~al.}{2023}]{BenAmi+2023PASP_LAST_Science}
{Ben-Ami} S.,  et~al., 2023, \mn@doi [arXiv e-prints]
  {10.48550/arXiv.2304.02719}, \href
  {https://ui.adsabs.harvard.edu/abs/2023arXiv230402719B} {p. arXiv:2304.02719}

\bibitem[\protect\citeauthoryear{{Britt} \& {Consolmagno}}{{Britt} \&
  {Consolmagno}}{2003}]{Britt+2003M&PS_StonyMeteroriteProperties}
{Britt} D.~T.,  {Consolmagno} G.~J.,  2003, \mn@doi [M\&PS]
  {10.1111/j.1945-5100.2003.tb00305.x}, \href
  {https://ui.adsabs.harvard.edu/abs/2003M&PS...38.1161B} {38, 1161}

\bibitem[\protect\citeauthoryear{{Brosch}, {Kaspi}, {Niv}  \&
  {Manulis}}{{Brosch}
  et~al.}{2015}]{Brosch+2015ApSS_C28_Telescope_WiseObservatory}
{Brosch} N.,  {Kaspi} S.,  {Niv} S.,   {Manulis} I.,  2015, \mn@doi [APSS]
  {10.1007/s10509-015-2500-7}, \href
  {https://ui.adsabs.harvard.edu/abs/2015Ap&SS.359...49B} {359, 9}

\bibitem[\protect\citeauthoryear{{Cheng} et~al.,}{{Cheng}
  et~al.}{2018}]{Cheng+2018P&SS_DART_mission_Didymos_DeflectionMission}
{Cheng} A.~F.,  et~al., 2018, \mn@doi [P\&SS] {10.1016/j.pss.2018.02.015},
  \href {https://ui.adsabs.harvard.edu/abs/2018P&SS..157..104C} {157, 104}

\bibitem[\protect\citeauthoryear{Cheng et~al.,}{Cheng
  et~al.}{2023}]{Cheng+2023Nature_DART_MomentumTransfer}
Cheng A.~F.,  et~al., 2023, \mn@doi [Nature] {10.1038/s41586-023-05878-z}

\bibitem[\protect\citeauthoryear{Daly et~al.,}{Daly
  et~al.}{2023}]{Daly+2023Nature_DART_KineticImpact}
Daly R.~T.,  et~al., 2023, \mn@doi [Nature] {10.1038/s41586-023-05810-5}

\bibitem[\protect\citeauthoryear{{Draine} \& {Lee}}{{Draine} \&
  {Lee}}{1984}]{Draine+1984ApJ_OpticalProperties_Graphite_Silicate_Grains}
{Draine} B.~T.,  {Lee} H.~M.,  1984, \mn@doi [\apj] {10.1086/162480}, \href
  {https://ui.adsabs.harvard.edu/abs/1984ApJ...285...89D} {285, 89}

\bibitem[\protect\citeauthoryear{{Fahnestock} et~al.,}{{Fahnestock}
  et~al.}{2022}]{Fahnestock+2022PSJ_DART_Predictions}
{Fahnestock} E.~G.,  et~al., 2022, \mn@doi [PSJ] {10.3847/PSJ/ac7fa1}, \href
  {https://ui.adsabs.harvard.edu/abs/2022PSJ.....3..206F} {3, 206}

\bibitem[\protect\citeauthoryear{{Farinella} \& {Davis}}{{Farinella} \&
  {Davis}}{1992}]{Farinella+Davis1992Icar_AsteroidsCollisionsVelocity}
{Farinella} P.,  {Davis} D.~R.,  1992, \mn@doi [Icarus]
  {10.1016/0019-1035(92)90060-K}, \href
  {https://ui.adsabs.harvard.edu/abs/1992Icar...97..111F} {97, 111}

\bibitem[\protect\citeauthoryear{{Gehrels}}{{Gehrels}}{1986}]{Gehrels1986_PoissonCI}
{Gehrels} N.,  1986, \mn@doi [\apj] {10.1086/164079}, \href
  {https://ui.adsabs.harvard.edu/abs/1986ApJ...303..336G} {303, 336}

\bibitem[\protect\citeauthoryear{{Gehrels} et~al.,}{{Gehrels}
  et~al.}{2004}]{Gehrels+2004_Swift}
{Gehrels} N.,  et~al., 2004, \mn@doi [\apj] {10.1086/422091}, \href
  {https://ui.adsabs.harvard.edu/abs/2004ApJ...611.1005G} {611, 1005}

\bibitem[\protect\citeauthoryear{Graykowski et~al.,}{Graykowski
  et~al.}{2023}]{Graykowski+2023Nature_DART_LightCurve}
Graykowski A.,  et~al., 2023, \mn@doi [Nature] {10.1038/s41586-023-05852-9}

\bibitem[\protect\citeauthoryear{{Gudebski}, {Heldridge}, {McGawn}, {Hill},
  {Swift}  \& {Zhou}}{{Gudebski}
  et~al.}{2023}]{Gudebski2023RNAAS_DART_Impact_Anomalous_Orbital_Decay}
{Gudebski} T.,  {Heldridge} E.,  {McGawn} B.,  {Hill} E.~O.,  {Swift} J.~J.,
  {Zhou} H.,  2023, \mn@doi [Research Notes of the American Astronomical
  Society] {10.3847/2515-5172/acf214}, \href
  {https://ui.adsabs.harvard.edu/abs/2023RNAAS...7..178G} {7, 178}

\bibitem[\protect\citeauthoryear{{Gustafson}}{{Gustafson}}{1994}]{Gustafson+1994AREPS_PhysicsZodiacalLight_Review}
{Gustafson} B.~A.~S.,  1994, \mn@doi [Annual Review of Earth and Planetary
  Sciences] {10.1146/annurev.ea.22.050194.003005}, \href
  {https://ui.adsabs.harvard.edu/abs/1994AREPS..22..553G} {22, 553}

\bibitem[\protect\citeauthoryear{{Haisch}, {Strong}  \& {Rodono}}{{Haisch}
  et~al.}{1991}]{Haisch+1991_FlareStars_Review}
{Haisch} B.,  {Strong} K.~T.,   {Rodono} M.,  1991, \mn@doi [\araa]
  {10.1146/annurev.aa.29.090191.001423}, \href
  {https://ui.adsabs.harvard.edu/abs/1991ARA&A..29..275H} {29, 275}

\bibitem[\protect\citeauthoryear{{Harker}, {Woodward}  \& {Wooden}}{{Harker}
  et~al.}{2005}]{Harker+2005Sci_DeepImpact_Tempel1_Dust_Collision}
{Harker} D.~E.,  {Woodward} C.~E.,   {Wooden} D.~H.,  2005, \mn@doi [Science]
  {10.1126/science.1119143}, \href
  {https://ui.adsabs.harvard.edu/abs/2005Sci...310..278H} {310, 278}

\bibitem[\protect\citeauthoryear{{Holsapple}}{{Holsapple}}{2022}]{Holsapple2022P&SS_MainBeltAsteroids_CollisionsHistory_Ejecta_Qstar}
{Holsapple} K.~A.,  2022, \mn@doi [P\&SS] {10.1016/j.pss.2022.105529}, \href
  {https://ui.adsabs.harvard.edu/abs/2022P&SS..21905529H} {219, 105529}

\bibitem[\protect\citeauthoryear{{Holsapple} \& {Housen}}{{Holsapple} \&
  {Housen}}{2019}]{Holsapple+Housen2019P&SS_AsteroidsCollisions_Review}
{Holsapple} K.~A.,  {Housen} K.~R.,  2019, \mn@doi [P\&SS]
  {10.1016/j.pss.2019.104724}, \href
  {https://ui.adsabs.harvard.edu/abs/2019P&SS..17904724H} {179, 104724}

\bibitem[\protect\citeauthoryear{{Housen}, {Schmidt}  \& {Holsapple}}{{Housen}
  et~al.}{1983}]{Housen+1983JGR_CrateringScalingLaws_DimensionalAnalysis}
{Housen} K.~R.,  {Schmidt} R.~M.,   {Holsapple} K.~A.,  1983, \mn@doi [\jgr]
  {10.1029/JB088iB03p02485}, \href
  {https://ui.adsabs.harvard.edu/abs/1983JGR....88.2485H} {88, 2485}

\bibitem[\protect\citeauthoryear{{Ishiguro} et~al.,}{{Ishiguro}
  et~al.}{2011}]{Masateru+2011ApJ_596_Scheila_AsteroidCollision_Observations}
{Ishiguro} M.,  et~al., 2011, \mn@doi [\apjl] {10.1088/2041-8205/740/1/L11},
  \href {https://ui.adsabs.harvard.edu/abs/2011ApJ...740L..11I} {740, L11}

\bibitem[\protect\citeauthoryear{{Jewitt}, {Weaver}, {Agarwal}, {Mutchler}  \&
  {Drahus}}{{Jewitt} et~al.}{2010}]{Jewitt+2010Natur_AsteroidCollision_P2010A2}
{Jewitt} D.,  {Weaver} H.,  {Agarwal} J.,  {Mutchler} M.,   {Drahus} M.,  2010,
  \mn@doi [Nature] {10.1038/nature09456}, \href
  {https://ui.adsabs.harvard.edu/abs/2010Natur.467..817J} {467, 817}

\bibitem[\protect\citeauthoryear{{Jewitt}, {Weaver}, {Mutchler}, {Larson}  \&
  {Agarwal}}{{Jewitt}
  et~al.}{2011}]{Jewitt+2011ApJ_596_Scheila_AsteroidCollision_HST_observations}
{Jewitt} D.,  {Weaver} H.,  {Mutchler} M.,  {Larson} S.,   {Agarwal} J.,  2011,
  \mn@doi [\apjl] {10.1088/2041-8205/733/1/L4}, \href
  {https://ui.adsabs.harvard.edu/abs/2011ApJ...733L...4J} {733, L4}

\bibitem[\protect\citeauthoryear{{Jewitt}, {Kim}, {Li}  \& {Mutchler}}{{Jewitt}
  et~al.}{2023}]{Jewitt+2023ApJ_DART_Impact_DimorphosBoulderSwarm}
{Jewitt} D.,  {Kim} Y.,  {Li} J.,   {Mutchler} M.,  2023, \mn@doi [\apjl]
  {10.3847/2041-8213/ace1ec}, \href
  {https://ui.adsabs.harvard.edu/abs/2023ApJ...952L..12J} {952, L12}

\bibitem[\protect\citeauthoryear{{Keller} et~al.,}{{Keller}
  et~al.}{2007}]{Keller+2007PASA_SkyMapper}
{Keller} S.~C.,  et~al., 2007, \mn@doi [\pasa] {10.1071/AS07001}, \href
  {https://ui.adsabs.harvard.edu/abs/2007PASA...24....1K} {24, 1}

\bibitem[\protect\citeauthoryear{{Kim} \& {Jewitt}}{{Kim} \&
  {Jewitt}}{2023}]{Kim+Jewitt2023arXiv_DART_Impact_SIngleEjectionToExplainDoubleTail}
{Kim} Y.,  {Jewitt} D.,  2023, arXiv e-prints, \href
  {https://ui.adsabs.harvard.edu/abs/2023arXiv230915116K} {p. arXiv:2309.15116}

\bibitem[\protect\citeauthoryear{{Kumamoto} et~al.,}{{Kumamoto}
  et~al.}{2022}]{Kumamoto+2022arXiv_DART_Didymos_PredictionSimulations}
{Kumamoto} K.~M.,  et~al., 2022, arXiv e-prints, \href
  {https://ui.adsabs.harvard.edu/abs/2022arXiv220911876K} {p. arXiv:2209.11876}

\bibitem[\protect\citeauthoryear{Li et~al.,}{Li
  et~al.}{2023}]{Li+2023Nature_DART_EjectaActiveAsteroid}
Li J.-Y.,  et~al., 2023, \mn@doi [Nature] {10.1038/s41586-023-05811-4}

\bibitem[\protect\citeauthoryear{{Lierle}, {Schmidt}, {Baumgardner}, {Moore}
  \& {Lovett}}{{Lierle}
  et~al.}{2023}]{Lierle+2023arXiv_HighResSpectroscopy_Mercury_NaI}
{Lierle} P.,  {Schmidt} C.,  {Baumgardner} J.,  {Moore} L.,   {Lovett} E.,
  2023, \mn@doi [arXiv e-prints] {10.48550/arXiv.2307.16872}, \href
  {https://ui.adsabs.harvard.edu/abs/2023arXiv230716872L} {p. arXiv:2307.16872}

\bibitem[\protect\citeauthoryear{{Luther}, {Zhu}, {Collins}  \&
  {Wunnemann}}{{Luther}
  et~al.}{2018}]{Luther+2018M&PS_AsteroidsCollisions_DART_ImpactModeling}
{Luther} R.,  {Zhu} M.-H.,  {Collins} G.,   {Wunnemann} K.,  2018, Materorite
  \& Plan. Sci., \href {https://ui.adsabs.harvard.edu/abs/2018M&PS...53.1705L}
  {53, 1705}

\bibitem[\protect\citeauthoryear{{McKay}, {Heiken}, {Taylor}, {Clanton},
  {Morrison}  \& {Ladle}}{{McKay}
  et~al.}{1972}]{McKay+1972LPSC_LunarRegolithParticlesSizeDistribution}
{McKay} D.~S.,  {Heiken} G.~H.,  {Taylor} R.~M.,  {Clanton} U.~S.,  {Morrison}
  D.~A.,   {Ladle} G.~H.,  1972, Lunar and Planetary Science Conference
  Proceedings, \href {https://ui.adsabs.harvard.edu/abs/1972LPSC....3..983M}
  {3, 983}

\bibitem[\protect\citeauthoryear{{Meech} et~al.,}{{Meech}
  et~al.}{2005}]{Meech+2005Sci_DeepImpact_Tempel1_Collision_Observations}
{Meech} K.~J.,  et~al., 2005, \mn@doi [Science] {10.1126/science.1118978},
  \href {https://ui.adsabs.harvard.edu/abs/2005Sci...310..265M} {310, 265}

\bibitem[\protect\citeauthoryear{{Moreno} et~al.,}{{Moreno}
  et~al.}{2011}]{Moreno+2011ApJ_596_Sceila_AsteroidCollision_Observations}
{Moreno} F.,  et~al., 2011, \mn@doi [\apj] {10.1088/0004-637X/738/2/130}, \href
  {https://ui.adsabs.harvard.edu/abs/2011ApJ...738..130M} {738, 130}

\bibitem[\protect\citeauthoryear{{Moreno} et~al.,}{{Moreno}
  et~al.}{2023}]{Moreno+2023PSJ_DART_Impact_Ejecta_Observations_and_Models}
{Moreno} F.,  et~al., 2023, PSJ, \href
  {https://ui.adsabs.harvard.edu/abs/2023PSJ.....4..138M} {4, 138}

\bibitem[\protect\citeauthoryear{{Ofek}}{{Ofek}}{2014}]{Ofek2014_MAAT}
{Ofek} E.~O.,  2014, {MATLAB package for astronomy and astrophysics}
  (\mn@eprint {ascl} {1407.005})

\bibitem[\protect\citeauthoryear{{Ofek}}{{Ofek}}{2019}]{Ofek2019_Astrometry_Code}
{Ofek} E.~O.,  2019, \mn@doi [\pasp] {10.1088/1538-3873/ab04df}, \href
  {https://ui.adsabs.harvard.edu/abs/2019PASP..131e4504O} {131, 054504}

\bibitem[\protect\citeauthoryear{{Ofek}}{{Ofek}}{2023}]{Ofek+2023ASP_LAST_PipelineI}
{Ofek} E.~O. e.~a.,  2023, submitted.

\bibitem[\protect\citeauthoryear{{Ofek} \& {Ben-Ami}}{{Ofek} \&
  {Ben-Ami}}{2020}]{Ofek+BenAmi2020_Grasp_SkySurvrys_CostEffectivness}
{Ofek} E.~O.,  {Ben-Ami} S.,  2020, arXiv e-prints, \href
  {https://ui.adsabs.harvard.edu/abs/2020arXiv201104674O} {p. arXiv:2011.04674}

\bibitem[\protect\citeauthoryear{{Rivkin} et~al.,}{{Rivkin}
  et~al.}{2021}]{Rivkin+2021PSJ_DART_MissionRequirments}
{Rivkin} A.~S.,  et~al., 2021, \mn@doi [PSJ] {10.3847/PSJ/ac063e}, \href
  {https://ui.adsabs.harvard.edu/abs/2021PSJ.....2..173R} {2, 173}

\bibitem[\protect\citeauthoryear{{Roming} et~al.,}{{Roming}
  et~al.}{2005}]{Roming+2005SSRv_Swift_UVOT}
{Roming} P. W.~A.,  et~al., 2005, \mn@doi [SSRv] {10.1007/s11214-005-5095-4},
  \href {https://ui.adsabs.harvard.edu/abs/2005SSRv..120...95R} {120, 95}

\bibitem[\protect\citeauthoryear{{Roth} et~al.,}{{Roth}
  et~al.}{2023}]{Roth+2023arXiv_DART_Impact_ALMA_Observations}
{Roth} N.~X.,  et~al., 2023, \mn@doi [arXiv e-prints]
  {10.48550/arXiv.2306.05908}, \href
  {https://ui.adsabs.harvard.edu/abs/2023arXiv230605908R} {p. arXiv:2306.05908}

\bibitem[\protect\citeauthoryear{{S{\'a}nchez}, {Scheeres}  \&
  {Quillen}}{{S{\'a}nchez}
  et~al.}{2022}]{Sanchez+2022arXiv_DART_Didymos_ImpactSeismicWaves}
{S{\'a}nchez} P.,  {Scheeres} D.~J.,   {Quillen} A.~C.,  2022, arXiv e-prints,
  \href {https://ui.adsabs.harvard.edu/abs/2022arXiv220911353S} {p.
  arXiv:2209.11353}

\bibitem[\protect\citeauthoryear{{Shestakova}, {Serebryanskiy}  \&
  {Aimanova}}{{Shestakova}
  et~al.}{2023}]{Shestakova+2023Icar_DART_Impact_Spectroscopy_EmissionLines}
{Shestakova} L.,  {Serebryanskiy} A.,   {Aimanova} G.,  2023, \mn@doi [\icarus]
  {10.1016/j.icarus.2023.115595}, \href
  {https://ui.adsabs.harvard.edu/abs/2023Icar..40115595S} {401, 115595}

\bibitem[\protect\citeauthoryear{{Soumagnac} \& {Ofek}}{{Soumagnac} \&
  {Ofek}}{2018}]{Soumagnac+Ofek2018_catsHTM}
{Soumagnac} M.~T.,  {Ofek} E.~O.,  2018, \mn@doi [\pasp]
  {10.1088/1538-3873/aac410}, \href
  {https://ui.adsabs.harvard.edu/abs/2018PASP..130g5002S} {130, 075002}

\bibitem[\protect\citeauthoryear{{Tancredi}, {Liu}, {Campo-Bagatin}, {Moreno}
  \& {Dom{\'\i}nguez}}{{Tancredi}
  et~al.}{2022}]{Tancredi+2022arXiv_DART_Didymos_SeismicWavesDustLofting}
{Tancredi} G.,  {Liu} P.-Y.,  {Campo-Bagatin} A.,  {Moreno} F.,
  {Dom{\'\i}nguez} B.,  2022, arXiv e-prints, \href
  {https://ui.adsabs.harvard.edu/abs/2022arXiv220902805T} {p. arXiv:2209.02805}

\bibitem[\protect\citeauthoryear{Thomas et~al.,}{Thomas
  et~al.}{2023}]{Thomas+2023Nature_DART_OrbitalPeriodChange}
Thomas C.~A.,  et~al., 2023, \mn@doi [Nature] {10.1038/s41586-023-05805-2}

\bibitem[\protect\citeauthoryear{{Wolf} et~al.,}{{Wolf}
  et~al.}{2018}]{Wolf+2018PASA_SkyMapper_DR1}
{Wolf} C.,  et~al., 2018, \mn@doi [\pasa] {10.1017/pasa.2018.5}, \href
  {https://ui.adsabs.harvard.edu/abs/2018PASA...35...10W} {35, e010}

\bibitem[\protect\citeauthoryear{{Zackay} \& {Ofek}}{{Zackay} \&
  {Ofek}}{2017}]{Zackay+2017_CoadditionI}
{Zackay} B.,  {Ofek} E.~O.,  2017, \mn@doi [\apj]
  {10.3847/1538-4357/836/2/187}, \href
  {https://ui.adsabs.harvard.edu/abs/2017ApJ...836..187Z} {836, 187}

\bibitem[\protect\citeauthoryear{{Zhang}, {Battams}, {Ye}, {Knight}  \&
  {Schmidt}}{{Zhang} et~al.}{2023}]{Zhang+2023PSJ_Phaethon3200_NaI_emission}
{Zhang} Q.,  {Battams} K.,  {Ye} Q.,  {Knight} M.~M.,   {Schmidt} C.~A.,  2023,
  \mn@doi [PSJ] {10.3847/PSJ/acc866}, \href
  {https://ui.adsabs.harvard.edu/abs/2023PSJ.....4...70Z} {4, 70}

\makeatother
\end{thebibliography}

\bsp	
\label{lastpage}

\end{document}